\documentclass[12pt,review]{elsarticle}

\usepackage{lineno,hyperref}
\usepackage{color,soul}
\modulolinenumbers[5]

\newcommand{\bgea}{\begin{eqnarray}}
\newcommand{\enea}{\end{eqnarray}}

\begin{document}

\begin{frontmatter}

\title{AlM$_2$B$_2$ (M=Cr, Mn, Fe, Co, Ni): a group of nanolaminated materials}

\author[Uppsala]{Krisztina K\'adas\corref{mycorrespondingauthor}}
\cortext[mycorrespondingauthor]{Corresponding author}
\ead{Krisztina.Kadas@physics.uu.se}

\author[Uppsala]{Diana Iu\c{s}an}

\author[KTH]{Johan Hellsvik}

\author[Uppchem]{Johan Cedervall}

\author[Uppchem]{Pedro Berastegui}

\author[Uppchem]{Martin Sahlberg}

\author[Uppchem]{Ulf Jansson}

\author[Uppsala]{Olle Eriksson}

\address[Uppsala]{Division of Materials Theory, 
Department of Physics and Astronomy,
Uppsala University, Box 516, SE-751 20, Uppsala, Sweden}
\address[KTH]{Department of Materials and Nanophysics, School of Information and 
Communication Technology, Electrum 229, Royal Institute of Technology (KTH), 
SE-164 40 Kista, Sweden}
\address[Uppchem]{Department of Chemistry - The \AA ngstr\"om Laboratory,
Uppsala University, Box 538, 751 21 Uppsala, Sweden}

\begin{abstract}
Combining theory with experiments, we study
the phase stability, elastic properties, electronic structure and hardness
of layered ternary borides
AlCr$_2$B$_2$, AlMn$_2$B$_2$, AlFe$_2$B$_2$, AlCo$_2$B$_2$, and AlNi$_2$B$_2$.
We find that the first three borides of this series are stable phases,
while AlCo$_2$B$_2$ and AlNi$_2$B$_2$ are metastable.
We show that the elasticity increases in the boride series, and
predict that
AlCr$_2$B$_2$, AlMn$_2$B$_2$, and AlFe$_2$B$_2$ are more brittle, while
AlCo$_2$B$_2$ and AlNi$_2$B$_2$ are more ductile.
We propose that the elasticity of AlFe$_2$B$_2$ can be improved
by alloying it with cobalt or nickel, or a combination of them.
We present evidence 
that these ternary borides represent nanolaminated systems.
Based on SEM measurements, we demonstrate that they exhibit the
delamination phenomena, 
which leads to a reduced hardness compared to 
transition metal mono- and diborides.
We discuss the background of delamination by analyzing
chemical bonding and theoretical work of separation in these borides.
\end{abstract}


\end{frontmatter}

\section{Introduction}
\label{intro}

Compounds with a layered structure have a potential to 
act as nanolaminated materials with unique properties. 
The most well known nanolaminates are the so-called 
MAX phases \cite{barsoum2000}.
They are ternary metal carbides or nitrides of the general formula
M$_{n+1}$AX$_n$ with n=1, 2 or 3, where M is a transition metal,
A is a group A element, and X is either carbon or nitrogen.
MAX phases have unique chemical and physical properties such as 
high electrical and thermal conductivity, high thermal shock resistance 
and damage tolerance, machinability, high oxidation resistance, etc.  
Many MAX phases also exhibit special deformation properties 
characterized by basal slip, kink and shear band deformation, 
and delaminations of individual grains \cite{barsoum2000}.
These properties can be explained by the nanolaminated 
structure where MX slabs with strong M-X bonds 
are separated by A-layers with weaker M-A bonds. 

Another group of compounds, which potentially can exhibit a similar 
type of properties are metal borides. 
In fact, Telle and co-workers showed that W$_2$B$_5$ with 
alternating W and boron layers, has similar mechanical properties 
as MAX phases \cite{telle2006},
and that this phase also exhibits delamination phenomena. 
Furthermore, Ade and Hillebrecht recently proposed that
ternary borides, such as Cr$_2$AlB$_2$, Cr$_3$AlB$_4$, and Cr$_4$AlB$_6$,
exhibit nanolaminated structures. They called these ternary borides
MAB phases in analogy to the more well-known MAX phases \cite{ade2015}. 
Very recently, MoAlB was shown to represent a nanolaminated system
\cite{kota2016,lu2016}, 
and Lu and co-workers also measured the atomic structure of 
nanolaminated AlCr$_2$B$_2$ and AlFe$_2$B$_2$ \cite{lu2016}.

\begin{figure}[h!]
\begin{center}
\includegraphics[width=12cm]{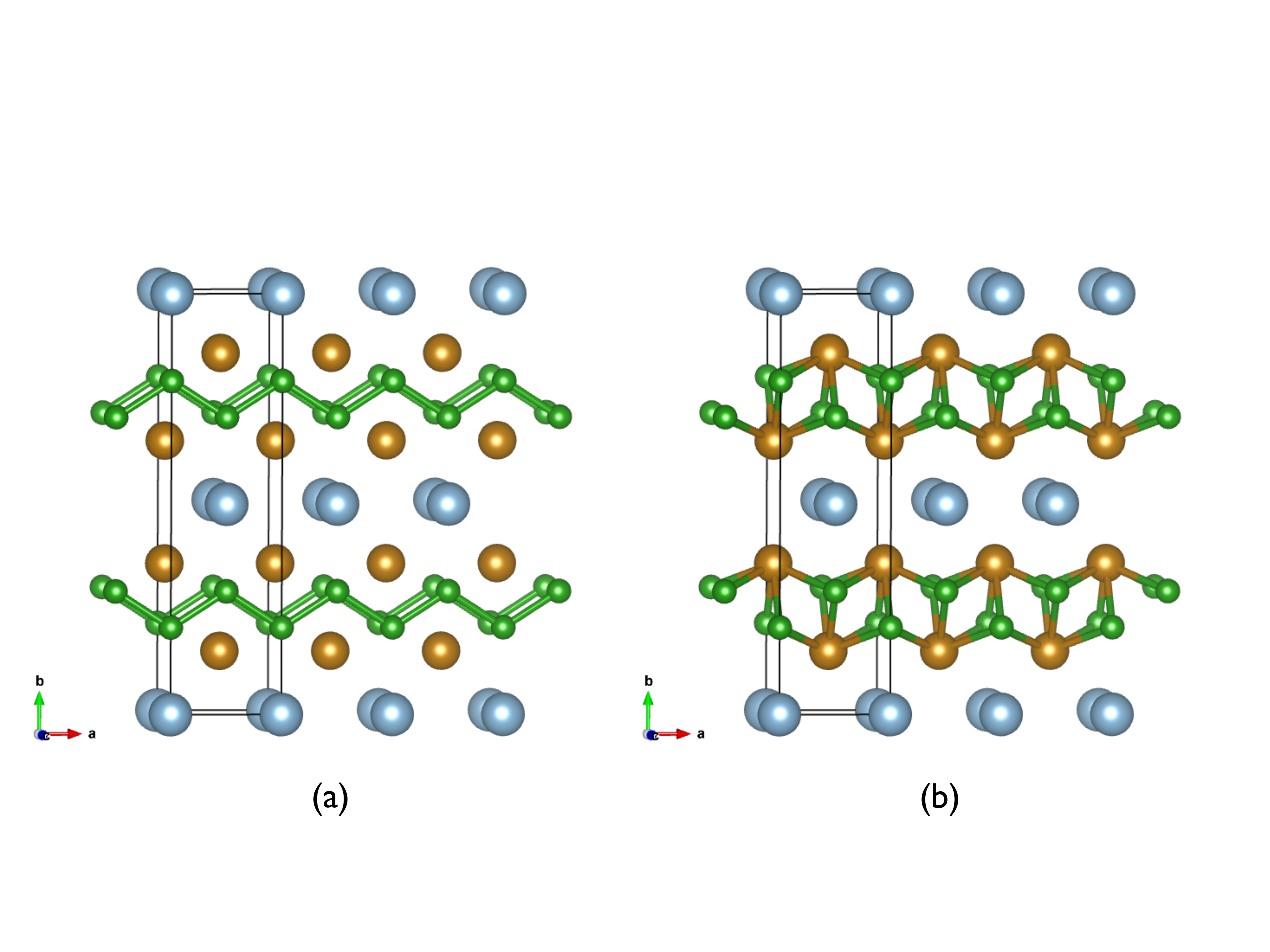}
\end{center}
\vskip -1.6cm
\caption{
(Color online.) 
Crystal structure of AlFe$_2$B$_2$.
The unit cell is shown by black lines.
Al, Fe and B atoms are displayed by blue, brown, and green, respectively.
The boron-boron bonds are shown on panel (a), while
the Fe-B bonds are displayed on panel (b).
}
\label{fig_str}
\end{figure}

The ternary AlM$_2$B$_2$ borides include in fact several known 
phases with M=Cr, Mn and Fe. 
The crystal structure of AlFe$_2$B$_2$ is shown in Fig. \ref{fig_str},
where the analogy to the MAX phases is clearly seen with Fe-B slabs 
separated by Al layers. 
There are strong B-B and Fe-B bonds in the boride slabs, 
which are separated with much weaker Fe-Al and Al-B bonds.
The interest in this group of ternary borides has increased after 
AlFe$_2$B$_2$ was
presented as a possible magnetocaloric material exhibiting 
a large magnetocaloric effect \cite{tan2013,cedervall2016}. 
Tan et al. found that AlFe$_2$B$_2$ is a soft ferromagnet with the ordering 
temperature of 282 K and 307 K, 
and a saturation magnetization of 1.515 $\mu_{\rm B}$ or 1.03 $\mu_{\rm B}$ 
per Fe atom, depending on the synthesis method. 
They also investigated other ternary borides, 
and reported the calculated magnetic properties and electronic structures of
AlMn$_2$B$_2$, AlCr$_2$B$_2$ and AlFe$_{2-x}$Mn$_x$B$_2$
\cite{chai2015}.
The authors of Ref. \cite{chai2015}
found that AlMn$_2$B$_2$ and AlCr$_2$B$_2$ are nonmagnetic,
and in AlFe$_{2-x}$Mn$_x$B$_2$
both the saturation magnetization and 
the ferromagnetic ordering temperature 
gradually decreases with increasing Mn content.
However, they did not study the phase stability of the ternary borides. 
The mechanical properties of ternary AlM$_2$B$_2$ borides are 
also not systematically described previously. 
Nie et al. calculated the elastic properties of AlCr$_2$B$_2$ \cite{nie2013},
while Cheng et al. published theoretical elastic constants for AlFe$_2$B$_2$
\cite{cheng2014}.
The elastic properties of the other ternary AlM$_2$B$_2$ borides, however, 
are not known.

The aim of this study is to present a systematic study of the 
AlM$_2$B$_2$ (M=Cr, Mn, Fe, Co, and Ni) phases using 
density functional theory (DFT) based methods
combined with experimental studies. 
We calculate the unit cell parameters of the borides, and compare them with 
those measured for experimentally synthesized samples. 
We calculate the phase stability, elastic properties 
and electronic structure
of these ternary 
borides, and analyze their chemical bonding. 
We also measure the hardness of the experimental samples, and 
characterize them by means of scanning electron microscopy (SEM).
Finally, we discuss the nanolaminated structure and behavior of these borides and 
their similarities to MAX-phases.

\section{Computational details}
\label{comp}

We performed first principle calculations by means
of the projector augmented wave \cite{blochl94,kresse99} method as
implemented in the Vienna ab initio simulation package
\cite{vasp1,vasp2,vasp3}.
This method is based on the 
density functional theory \cite{hohenberg64,kohn65}.
To calculate the exchange-correlation
energy, the generalized gradient
approximation with the Perdew, Burke, and Ernzerhof
functional \cite{pbe} was used.
We employed a plane-wave
energy cutoff of 500 eV and used a Monkhorst-Pack grid of
16x8x16 to sample the Brillouin zones for geometry optimizations,
and a larger grid 30x30x30 for elastic constant and 
electronic structure calculations.
The conjugate-gradient method was applied to relax the atoms into their 
optimal positions until the forces on all atoms were smaller than 
0.005 eV/\AA.
We calculated 
the nine independent elastic constants 
$c_{11}$, $c_{22}$, $c_{33}$, $c_{44}$, $c_{55}$, $c_{66}$, $c_{12}$, $c_{13}$ and $c_{23}$
as described in Ref. \cite{ravindran98}.
Accordingly, we employed small strains to the equilibrium lattice, 
and deduced the elastic constants from the calculated total energies
of the distorted lattices.
In addition, magnetic moments were also calculated using
the full-potential linear muffin-tin orbital (FP-LMTO) code RSPt
\cite{Andersen:1975kh, Wills:2010ej}.

In order to investigate the chemical bonding between the atomic 
constituents, we have carried out crystal orbital Hamilton population (COHP) 
calculations
using the LOBSTER code 
\cite{Dronskowski:1993jp, Deringer:2011fg, Maintz:2013cp, Maintz:2016ee}.
To determine work of separation ($W$) in AlM$_2$B$_2$ borides, we
calculated the energy cost of separating these crystals 
by 20 \AA \ vacuum
between different layers perpendicular to crystallographic axis $b$.
Upon separation, atoms of the first two layers at the interfaces 
were allowed to relax, while the rest of the atoms were kept fixed.

\section{Experimental}
\label{exp}

Samples were synthesized by arc-melting stoichiometric amounts of chromium
(Alfa Aesar, purity 99.995\%), manganese (Institute of Physics, Polish
Academy of Sciences, purity 99.999\%), iron (Leico Industries, purity 99.995\%.
Surface oxides were reduced in H$_2$-gas.),
cobalt (Johnson Matthey, purity 99,999\%) or
nickel (ESPI Metals, purity 99,995\%),
with boron (Wacher-Chemie, purity 99.995\%) to their respective metal-boride.
The metal borides were then reacted with aluminum
(Gr\"anges SM, purity 99.999\%)
with an excess of 50\% aluminum \cite{elmassalami2011}
to suppress the formation of
secondary phases. All samples were then crushed, pressed into pellets
and heat treated in evacuated silica tubes at 900$^{\circ}$C for 14 days.
The chromium, manganese and iron samples were then etched in
diluted HCl (1:1) to remove impurity phases.

The crystalline phases were analysed with X-ray powder diffraction (XRD)
on a Bruker D8 Advance using CuK$\alpha$ radiation.
To precisely determine the unit cell parameters refinements were
performed in the program UnitCell \cite{holland1997},
this was done after the etching step for the chromium,
manganese and iron samples and directly after arc-melting
for the cobalt and nickel samples.

The heat treated pieces were remelted, annealed and placed
placed in bakelite and polished for 
Vickers micro-hardness measurements, which were performed on a 
Matsuzawa MTX50 with a load of 200 g dwelling for 15 s. 
The measurements were done 10 times in the same region on each sample 
and the mean value of the hardness values is reported here. 
Delamination of the samples were studied with 
scanning electron microscopy (SEM) using a Zeiss LEO 1550 equipped 
with an Aztec energy dispersive X-ray spectrometer (EDS). 
The polished samples were damaged by pressing a 
sharp diamond tip into the polished sample surface.

\section{Results}
\label{res}

\subsection{Crystal structure}
\label{str}

Ternary AlM$_2$B$_2$ borides (M=Cr, Mn, Fe) crystallize 
in an orthorhombic lattice, space group $Cmmm$.
The crystal structure of AlFe$_2$B$_2$ was first described by Jeitschko
in 1969 \cite{jeitschko69}.
These borides have a layered structure, where M-B layers are
separated by Al layers. 
The M-B layers consist of two boron layers between two 
transition metal layers.
In Fig. \ref{fig_str} we show the crystal structure of AlFe$_2$B$_2$.
The Fe-B layers include zigzag chains of boron atoms along the 
crystallographic axis $a$, which is displayed in Fig. \ref{fig_str}a.
The 
B-B bond length in these chains is 1.74 \AA.
In addition, Fe-B bonds are also formed in the Fe-B layers,
as shown in Fig. \ref{fig_str}b.
Each Fe atom forms six Fe-B bonds with a bond length of 2.16 \AA \ and
2.17 \AA, where the shorter bonds are between Fe atoms and B atoms of the
boron layer closer to the Fe layer, 
while the longer Fe-B bonds are formed between Fe atoms and
B atoms of the adjacent boron layer.
Each B atom is coordinated by six Fe atoms in the Fe-B layers.
In the Al layers between the Fe-B layers, each Al atom forms
Fe-Al bonds with Fe atoms of both neighboring Fe-B layers,
with a bond length of 2.61 \AA.
Al atoms also form bonds with two axial B atoms of the neighboring Fe-B layers,
the Al-B bond lengths being 2.29 \AA.

The theoretical and experimental unit cell parameters are listed in
Table \ref{tab_theor_cellpar}.
Theory reproduces the experimental cell parameters well,
the maximal deviation being 0.5\% in $a$,
0.1\% in $b$, and 2.4\% in lattice parameter $c$.
We note that our measured lattice parameters are very close to those measured
by Ade and Hillebrecht \cite{ade2015} the difference being $\leq$0.3\%,
and also to those measured by Chai et al. \cite{chai2015}, 
where the maximal deviation is 0.8\%.
For the first three borides, namely for AlCr$_2$B$_2$, AlMn$_2$B$_2$, and AlFe$_2$B$_2$,
where measured lattice parameters are available,
the experimental volume of the unit cell decreases in the series 
from AlCr$_2$B$_2$ to AlFe$_2$B$_2$ 
according to both our experimental results and 
those published by Ade and Hillebrecht \cite{ade2015}, and Chai et al. \cite{chai2015}.
In the calculations, however, AlMn$_2$B$_2$ has the lowest volume 
of these three borides,
which is due to the fact that its theoretical lattice parameter $c$ 
is smaller than the experimental one by 2.4\%.

\subsection{Phase stability}
\label{stab}

To examine the stability of AlM$_2$B$_2$ (M=Cr, Mn, Fe, Co, Ni),
we calculate the formation energies ($\Delta E$)
from competing binary phases:
\begin{equation}
{\rm AlB}_2 + 2 {\rm CrB} + {\rm Cr_2Al}
\rightarrow
2 {\rm AlCr_2B_2},
\label{eq_Cr}
\end{equation}

\begin{equation}
{\rm AlB}_2 + 24 {\rm MnB} + 2 {\rm MnAl_6}
\rightarrow
13 {\rm AlMn_2B_2},
\label{eq_Mn}
\end{equation}

\begin{equation}
{\rm AlB}_2 + 4 {\rm FeB} + 2 {\rm FeAl}
\rightarrow
3 {\rm AlFe_2B_2},
\label{eq_Fe}
\end{equation}

\begin{equation}
{\rm AlB}_2 + 4 {\rm CoB} + 2 {\rm CoAl}
\rightarrow
3 {\rm AlCo_2B_2},
\label{eq_Co}
\end{equation}

\begin{equation}
{\rm AlB}_2 + 4 {\rm NiB} + 2 {\rm NiAl}
\rightarrow
3 {\rm AlNi_2B_2},
\label{eq_Ni}
\end{equation}
where
AlB$_2$ \cite{felten56} is of $P6/mmm$ structure,
MnB \cite{kiessling50},
FeB \cite{bjurstrom33} and CoB \cite{bjurstrom33} are of $Pnma$ structure,
CrB \cite{okada87} and NiB \cite{lugscheider74} is of $Cmcm$ structure,
Cr$_2$Al \cite{kallel69} is of $I$4/$mmm$ structure,
MnAl$_6$ \cite{kontio81} is of $Cmcm$ structure,
and finally,
FeAl \cite{vanderkraan86}, CoAl \cite{ipser92} and NiAl \cite{dutchak81}
are of $Pm\bar{3}m$ (CsCl) structure.
A negative formation energy means that AlM$_2$B$_2$ 
represent a stable phase.

\begin{figure}[h!]
\begin{center}
\includegraphics[width=10cm]{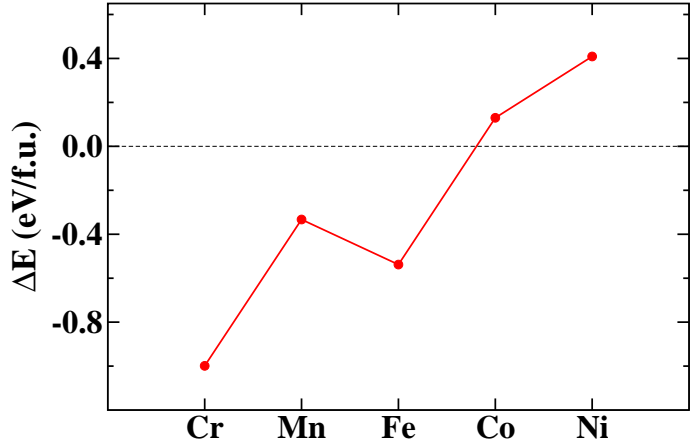}
\end{center}
\vskip -0.6cm
\caption{
(Color online.)
Theoretical formation energies of AlM$_2$B$_2$ (M=Cr, Mn, Fe, Co, Ni).
}
\label{fig_deltae}
\end{figure}

\begin{figure}[h!]
\begin{center}
\includegraphics[width=10cm]{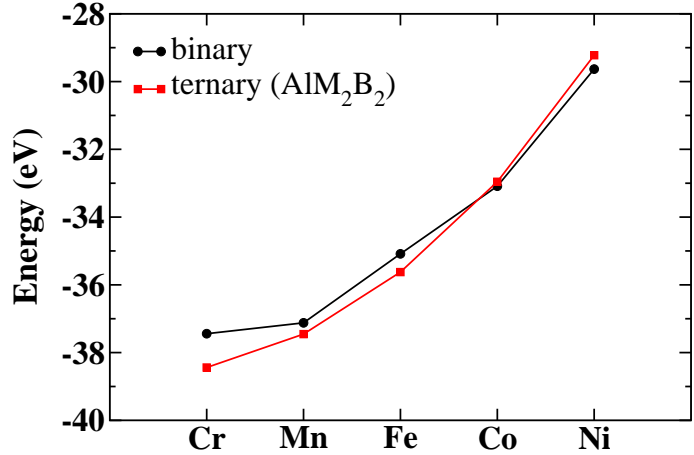}
\end{center}
\vskip -0.6cm
\caption{
(Color online.)
Total energies of the binary phases in reactions \ref{eq_Cr}-\ref{eq_Ni}
(black circles) and the ternary AlM$_2$B$_2$ phases (red squares).
}
\label{fig_energies}
\end{figure}

We calculate 
$\Delta E_1$=-0.999 eV/f.u., 
$\Delta E_2$=-0.333 eV/f.u., 
$\Delta E_3$=-0.538 eV/f.u. formation energy for 
reaction \ref{eq_Cr}, \ref{eq_Mn}, and \ref{eq_Fe}, respectively,
i.e. AlCr$_2$B$_2$, AlMn$_2$B$_2$, and AlFe$_2$B$_2$ are stable
(see Fig. \ref{fig_deltae}).
For reactions \ref{eq_Co} and \ref{eq_Ni}, however,
we obtain positive formation energies
(Fig. \ref{fig_deltae}).
We calculate
$\Delta E_4$=0.130 eV/f.u. for the formation of AlCo$_2$B$_2$
(reaction \ref{eq_Co}), and 
$\Delta E_5$=0.409 eV/f.u. for the formation of AlNi$_2$B$_2$ 
(reaction \ref{eq_Ni}).
Our results suggest that both AlCo$_2$B$_2$ and AlNi$_2$B$_2$ are 
thermodynamically unstable, and at best could be formed as
metastable compounds.
However,
the calculated phase stabilities (Fig. \ref{fig_deltae}) suggest
that AlCo$_2$B$_2$ and AlNi$_2$B$_2$ alloyed with Fe can be stable phases.

In Fig. \ref{fig_energies} we plot the calculated total energies of the binary
phases in reactions \ref{eq_Cr}-\ref{eq_Ni} corresponding to one formula unit of
AlM$_2$B$_2$ (left-hand sides in Eqs. \ref{eq_Cr}-\ref{eq_Ni}), 
together with the energies of the ternary AlM$_2$B$_2$ phases 
(right-hand sides in Eqs. \ref{eq_Cr}-\ref{eq_Ni}).
Figure \ref{fig_energies} shows that the energy of the ternary phases increases
in the series, i.e.  from AlCr$_2$B$_2$ to AlNi$_2$B$_2$,
and the stability of AlCr$_2$B$_2$, AlMn$_2$B$_2$ and AlFe$_2$B$_2$ 
is due to the high energy of the corresponding binary phases.

\subsection{Elastic properties}
\label{elastic}

The calculated single crystal elastic constants $c_{\rm ij}$ 
are shown in Table \ref{tab_cij}.
The elastic constant $c_{\rm 11}$ is larger than 
the other two principal elastic constants
$c_{\rm 22}$ and $c_{\rm 33}$
in all ternary borides. 
This means that the AlM$_2$B$_2$ crystals 
are harder to compress along axis $a$, than along axes $b$ or $c$,
which is in line with the fact that the strong boron-boron bonds can be found
along axis $a$ in these crystals.
We note that Cheng et al. \cite{cheng2014} found
$c_{\rm 22}>c_{\rm 11}>c_{\rm 33}$ in AlFe$_2$B$_2$, 
in contrast to our results, 
which could be due to a smaller number of 
k-points that they applied to calculate the elastic constants,
compared to what we used.
We find that the shear elastic constants are significantly smaller
than the principal ones in all AlM$_2$B$_2$ borides, see Table \ref{tab_cij}.
Our theoretical elastic constants for AlCr$_2$B$_2$ agree with those
calculated by Nie et al. \cite{nie2013}, the deviation being less than 3\%
in the principal elastic constants, and less than 10\% in the shear elastic 
constants. 
The difference may again be due to the smaller k-mesh employed in Ref. \cite{nie2013}.

For an orthorhombic crystal, the criteria for mechanical stability are
$c_{11}>0$,
$c_{11}+c_{22}>c_{22}^2$, 
$c_{11}c_{22}c_{33}+2c_{12}c_{13}c_{23}-c_{11}c_{23}^2-c_{22}c_{13}^2-c_{33}c_{12}^2>0$,
$c_{44}>0$,
$c_{55}>0$,
$c_{66}>0$ \cite{mouhat2014}.
All of these stability criteria are fulfilled
in all the ternary borides examined in this paper, which means that 
AlCr$_2$B$_2$, AlMn$_2$B$_2$, AlFe$_2$B$_2$, AlCo$_2$B$_2$, and AlNi$_2$B$_2$
are mechanically stable.
Note that this is not equivalent to the phase stability 
discussed above.

\begin{figure}[h!]
\begin{center}
\includegraphics[width=10cm]{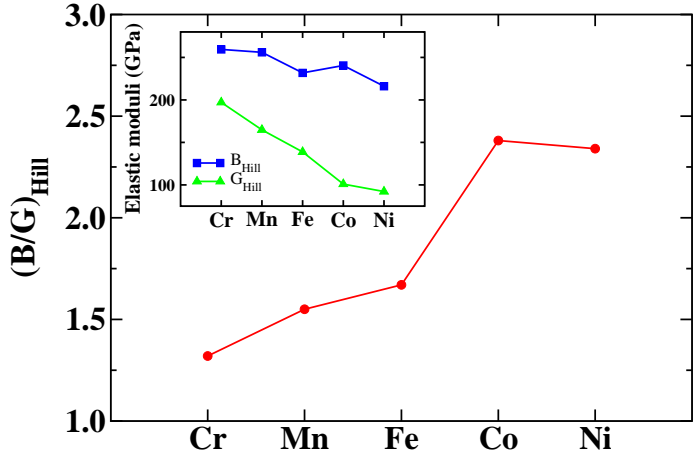}
\end{center}
\vskip -0.6cm
\caption{
(Color online.)
Theoretical $B/G$ (Hill averages) of AlM$_2$B$_2$.
Inset: calculated bulk moduli ($B$) and shear moduli ($G$)
of AlM$_2$B$_2$.
}
\label{fig_bg}
\end{figure}

From the single crystal elastic constants and the 
elastic compliance constants $s_{\rm ij}$,
we also calculated the polycrystalline elastic constants, 
namely the bulk moduli ($B$) and shear moduli ($G$) 
according to the Voigt ($B_{\rm V}$, $G_{\rm V}$) \cite{voigt} and 
Reuss approximations ($B_{\rm R}$, $G_{\rm R}$) \cite{reuss},
which correspond to the
upper and lower limit of the elastic moduli.
Hill's bulk ($B_{\rm H}$) and shear moduli ($G_{\rm H}$)
are the average of the Voigt and Reuss bounds, i.e. 
$B_{\rm H}=\frac{1}{2}(B_{\rm R}+B_{\rm V})$, and 
$G_{\rm H}=\frac{1}{2}(G_{\rm R}+G_{\rm V})$.
The Young's modulus ($E$) and Poisson ratio ($\nu$)
for an isotropic material is then
\begin{equation}
E_{\rm V,R,H} = \frac{9 B_{\rm V,R,H} G_{\rm V,R,H}}{3 B_{\rm V,R,H} + G_{\rm V,R,H}},
\label{eq_youngmod}
\end{equation}
and
\begin{equation}
\nu_{\rm V,R,H} = \frac{3 B_{\rm V,R,H} - 2 G_{\rm V,R,H}}{2(3 B_{\rm V,R,H} + G_{\rm V,R,H})}.
\label{eq_poisson}
\end{equation}
The calculated polycrystalline elastic constants are listed 
in Table \ref{tab_polycr_elastic_const}.
The bulk modulus to shear modulus ratio, $B/G$, 
is an important measure of elasticity of a material,
it can be employed to characterize the deformation behavior of a crystal.
Materials with a $B/G$$<$1.75 are expected to be brittle, 
while materials with $B/G$$>$1.75 are ductile.
Analyzing the bulk and shear moduli, we find that
both $B$ and $G$ decrease in the series, i.e.
from AlCr$_2$B$_2$ to AlNi$_2$B$_2$,
but the shear modulus decreases to a larger extent 
(see inset in Fig. \ref{fig_bg}).
While $B_{\rm H}$ decreases with 16.7\%, comparing AlNi$_2$B$_2$ to AlCr$_2$B$_2$,
we obtain a 53.2\% decrease for the shear modulus $G_{\rm H}$.
Accordingly, $B/G$ increases in the series,
except for AlNi$_2$B$_2$, which has a slightly smaller $B/G$ than AlCo$_2$B$_2$
(see Fig. \ref{fig_bg} and Table \ref{tab_polycr_elastic_const}).
The relatively high $B/G$ calculated for AlCo$_2$B$_2$ is due to its low
shear modulus and high bulk modulus.
Based on the theoretical $B/G$ ratios, we expect that 
AlCr$_2$B$_2$, AlMn$_2$B$_2$, and AlFe$_2$B$_2$ are more brittle, while
AlCo$_2$B$_2$ and AlNi$_2$B$_2$ are more ductile.
The relatively low $B/G$ calculated for AlFe$_2$B$_2$ is due to its
low bulk modulus (see inset in Fig. \ref{fig_bg}).
Based on our results, the elasticity of AlFe$_2$B$_2$,
which is the last stable boride of the series, 
can be improved, in fact it should be possible to be made ductile 
by alloying it with Co, or Ni, or a combination of them.

The Young's modulus, $E$, characterizes the bond strength of materials.
We obtain a decreasing trend in the theoretical Young's moduli 
from AlCr$_2$B$_2$ to AlNi$_2$B$_2$. 
We will discuss bond strengths in detail in the following section.

In general, the Poisson ratio is a measure of the stability of a crystal against 
shear.
Our calculated Poisson ratio follows an increasing trend 
from AlCr$_2$B$_2$ to AlCo$_2$B$_2$, and AlNi$_2$B$_2$ has a slightly lower
$\nu$ than AlCo$_2$B$_2$ (Table \ref{tab_polycr_elastic_const}).
Accordingly, the stability against shear decreases in the series.

The shift from brittle to ductile behavior from AlCr$_2$B$_2$ to AlNi$_2$B$_2$
finds further support from the analysis of Cauchy pressures.
In orthorhombic materials, the Cauchy pressure can be defined 
for the three different directions:
$P_{\rm a}$=$c_{23}$-$c_{44}$,
$P_{\rm b}$=$c_{13}$-$c_{55}$, and
$P_{\rm c}$=$c_{12}$-$c_{66}$.
In general, a positive Cauchy pressure is an indication of ductile behavior.
In AlCr$_2$B$_2$, the Cauchy pressure is negative for all the three directions, 
showing that this material is more brittle. 
In AlMn$_2$B$_2$, $P_{\rm c}$ one of the Cauchy pressures, 
namely $P_{\rm c}$, is positive. 
In the next boride of the series, AlFe$_2$B$_2$,
we obtain positive Cauchy pressure in two directions, here
$P_{\rm a}>$0 and $P_{\rm c}>$0.
In the last two borides, AlCo$_2$B$_2$ and AlNi$_2$B$_2$, the Cauchy pressure
is positive in all the three directions.
This is in line with the increasing trend of the $B/G$ ratio we showed
above.

It is established that elastic anisotropy plays an important role in
the formation of microcracks in ceramics \cite{tvergaard88}.
Hence, to contribute to the understanding of the mechanical properties of
AlM$_2$B$_2$ ternary borides, 
in the following we will analyze their anisotropy.
The shear anisotropic factors measure the degree of anisotropy 
in the bonding between atoms in different planes.
The shear anisotropic factor for the \{100\} shear planes between
the $<011>$ and $<010>$ directions is \cite{ravindran98}
\begin{equation}
A_1 = \frac{4c_{44}}{c_{11}+c_{33}-2c_{13}},
\label{eq_a1}
\end{equation}
for the \{010\} shear planes between the $<101>$ and $<001>$ directions it is
\begin{equation}
A_2 = \frac{4c_{55}}{c_{22}+c_{33}-2c_{23}},
\label{eq_a2}
\end{equation}
and finally for the \{001\} shear planes between the $<110>$ and $<010>$ directions it is
\begin{equation}
A_3 = \frac{4c_{66}}{c_{11}+c_{22}-2c_{12}}.
\label{eq_a3}
\end{equation}
For an isotropic crystal $A_i$=1, where $i$=1-3.
For anisotropic crystals $A_i$ can be larger or smaller than one, the 
difference being a measure of the degree of elastic anisotropy of the 
crystal.
Our calculated shear anisotropic factors are shown in Table \ref{tab_anisotr}.
We find that all AlM$_2$B$_2$ ternary borides are elastically anisotropic.
In most borides we obtain the highest values for the \{010\} shear planes
($A_2$),
except in AlMn$_2$B$_2$, where $A_3$ is the highest.

In Table \ref{tab_anisotr} we also list the percentage anisotropy in
compressibility ($A_B$=($B_{\rm V}$-$B_{\rm R}$)/($B_{\rm V}$+$B_{\rm R}$), and
in shear ($A_G$=($G_{\rm V}$-$G_{\rm R}$)/($G_{\rm V}$+$G_{\rm R}$), 
introduced by Chung and Buessem \cite{chung}.
$A_B$ and $A_G$ varies between 0 and 100\%, zero representing
an isotropic crystal.
The highest anisotropy in compressibility is around one percent,
we calculate $A_B$=1.09\% in AlNi$_2$B$_2$, and 
$A_B$=0.93\% in AlFe$_2$B$_2$.
We obtain higher anisotropies in shear, and we
find that $A_G$ increases in the boride series, 
from AlCr$_2$B$_2$ to AlNi$_2$B$_2$. 
Furthermore,
AlCo$_2$B$_2$ and AlNi$_2$B$_2$ have significantly higher anisotropies
than the first three borides, 
we calculate
$A_G$=4.72\% in AlCo$_2$B$_2$, and $A_G$=4.82\% in AlNi$_2$B$_2$.
Finally, we also calculated the universal anisotropy index,
$A_{\rm u}=5\frac{G_{\rm V}}{G_{\rm R}}+\frac{B_{\rm V}}{B_{\rm R}}-6$,
introduced by Ranganathan and Ostoja-Starzewski \cite{ranganathan2008}.
$A_{\rm u}$ can be positive or zero,
zero representing an isotropic crystal,
and accounts for both compressibility and shear contributions.
$A_{\rm u}$ changes similarly to $A_G$ in the boride series, 
i.e. it increases from AlCr$_2$B$_2$ to AlNi$_2$B$_2$, 
with AlCo$_2$B$_2$ and AlNi$_2$B$_2$ having notably higher anisotropies 
(0.50 and 0.53, respectively)
than the other three borides.

To estimate the hardness of AlM$_2$B$_2$ ternary borides, 
we can apply different macroscopic models for hardness prediction 
according to Ivanovskii \cite{ivanovskii2013}. 
These models represent semiempirical correlations between the 
Vickers hardness $H_{\rm V}$, and bulk, shear and Young moduli. 
All models show a general trend where AlCr$_2$B$_2$ should be 
the hardest phase followed by a decrease in hardness as the atomic 
number of M increases with AlNi$_2$B$_2$ as the least hard compound.  
Irrespective of the model, all thermodynamic stable borides 
(M=Cr, Mn and Fe) should exhibit hardness values above 17 GPa. 
For example, using the relation:
$H_{\rm V} = 0.1475 G$,
the hardness can be estimated  
29.1 GPa (AlCr$_2$B$_2$), 24.3 GPa (AlMn$_2$B$_2$), 20.5 GPa (AlFe$_2$B$_2$), 
14.9 GPa (AlCo$_2$B$_2$), and 13.6 GPa (AlNi$_2$B$_2$).

\subsection{Electronic structure and chemical bonding}
\label{dos}

\begin{figure}[h!]
\begin{center}
\includegraphics[width=0.45\textwidth]{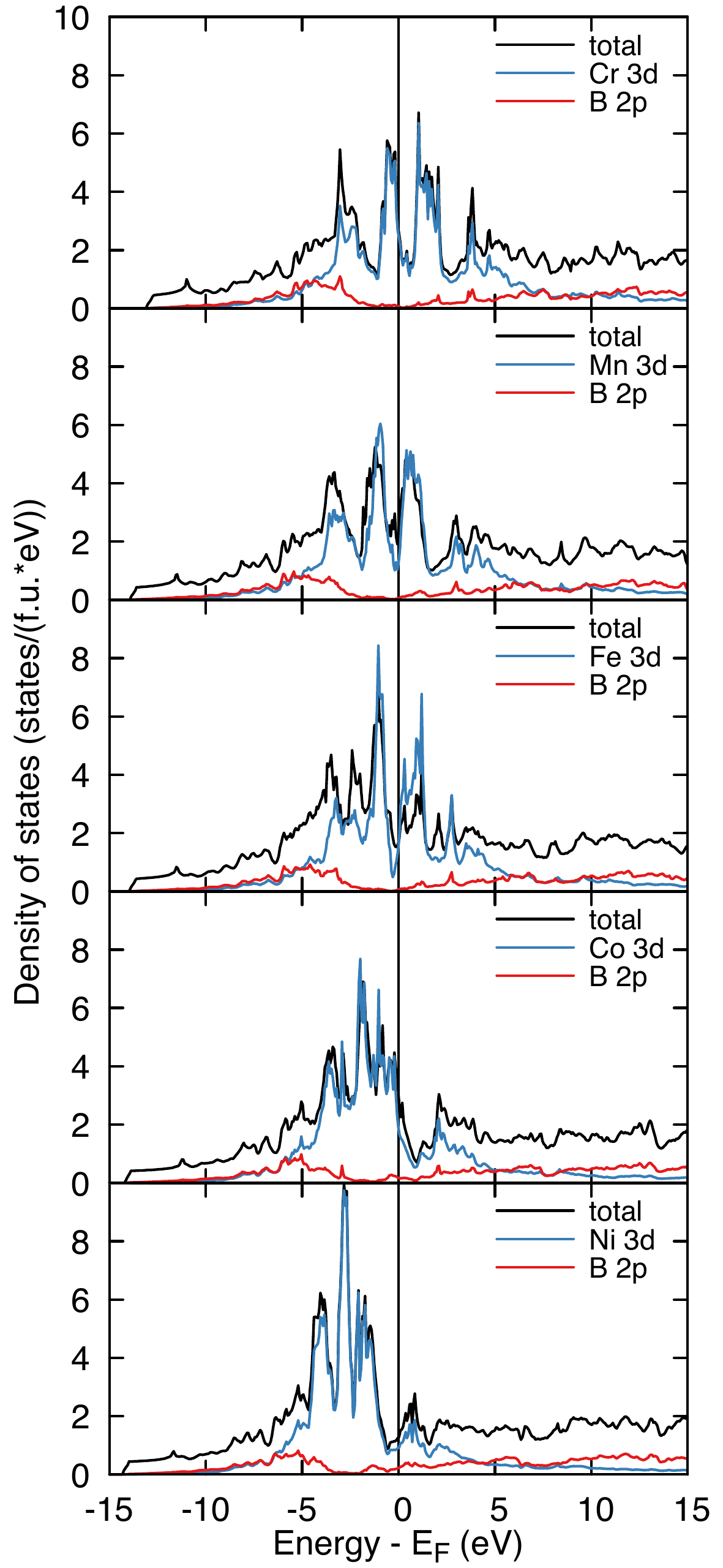}
\end{center}
\caption{\label{fig:dos-tm2alb2} 
(Color online.)
Density of states of the AlM$_2$B$_2$ compounds (M = Cr, Mn, Fe, Co, Ni).}
\end{figure}

Fig. \ref{fig:dos-tm2alb2} shows the density of states of the 
AlM$_2$B$_2$ compounds, with M ranging from Cr to Ni, 
as calculated with the FP-LMTO (RSPt) code. 
For a few eV below the Fermi level, the valence band has 
predominantly {\textit 3d} character, while the lower lying states are 
dominated by B and Al. 
As the atomic number of the transition metal, $Z$, increases, 
the M {\textit 3d} states and consequently the 
B {\textit 2p} are pushed to lower energy. 
The width of the {\textit 3d} decreases as well as their 
occupation increases, which leads to a decrease of the crystal field 
splitting of the transition metal states with increasing $Z$.

The calculated magnetic moments of the AlM$_2$B$_2$ compounds are 
shown in Table \ref{tab:magmom1}. 
A good agreement is obtained between the FP-LMTO and PAW results, 
with the exception of AlCo$_2$B$_2$ for which a small but finite moment is found 
within the former method. 
The energy difference between a spin-degenerate solution and a 
spin-polarized solution, with a small net moment is sometimes very small, 
and this small difference can approach the tine energy differences 
provided by different electronic structure methods (of order 1 meV/atom)
\cite{lejaeghere2016}. 
Unfortunately there are no experimental values with which to compare 
our calculated magnetic properties for AlCo$_2$B$_2$.
The induced moments onto the Al and B sites are negligible. 
The AlCr$_2$B$_2$ and AlNi$_2$B$_2$ compounds are found to be nonmagnetic.

\begin{figure}[h!]
\begin{center}
(a)\includegraphics[width=0.45\textwidth]{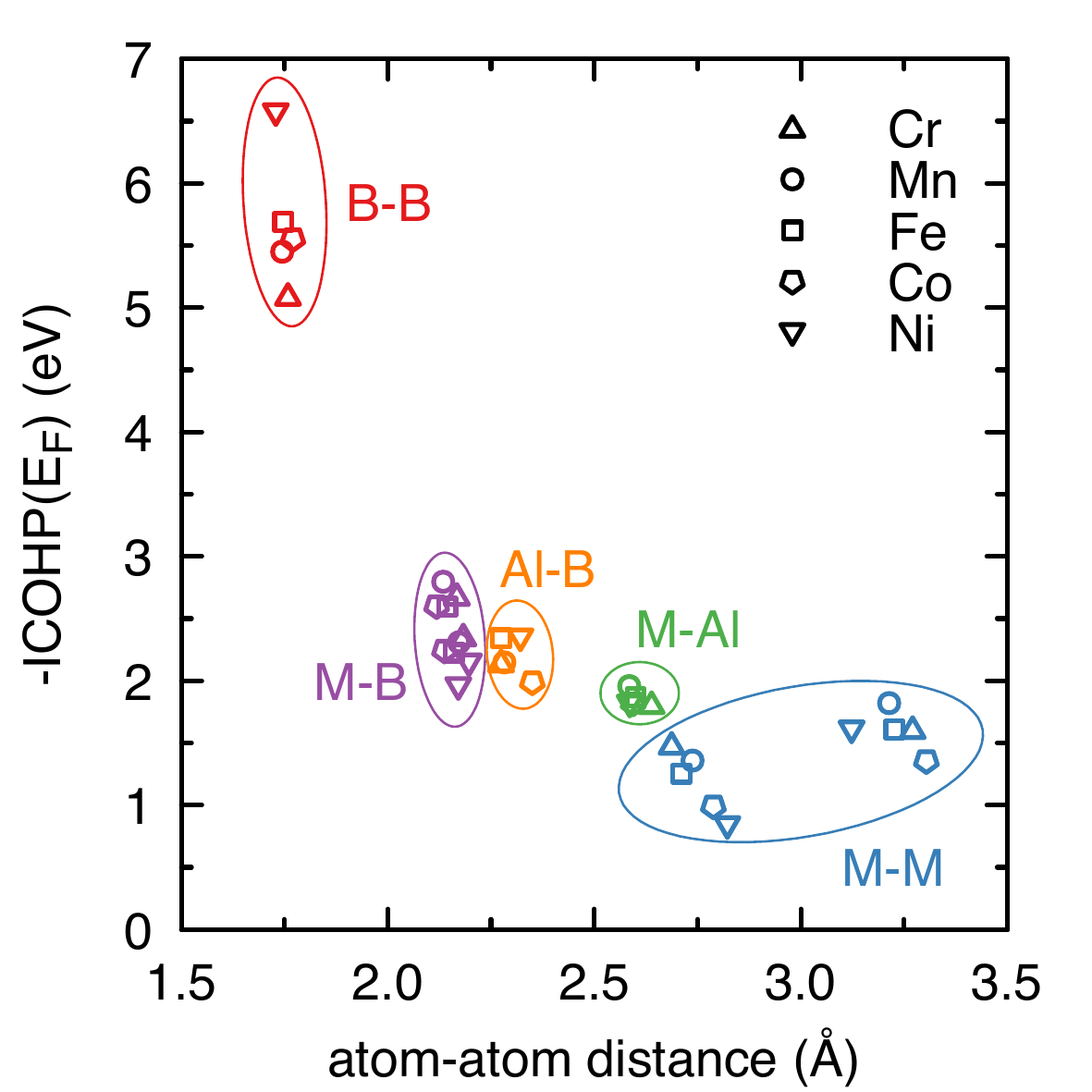} \\
(b)\includegraphics[width=0.45\textwidth]{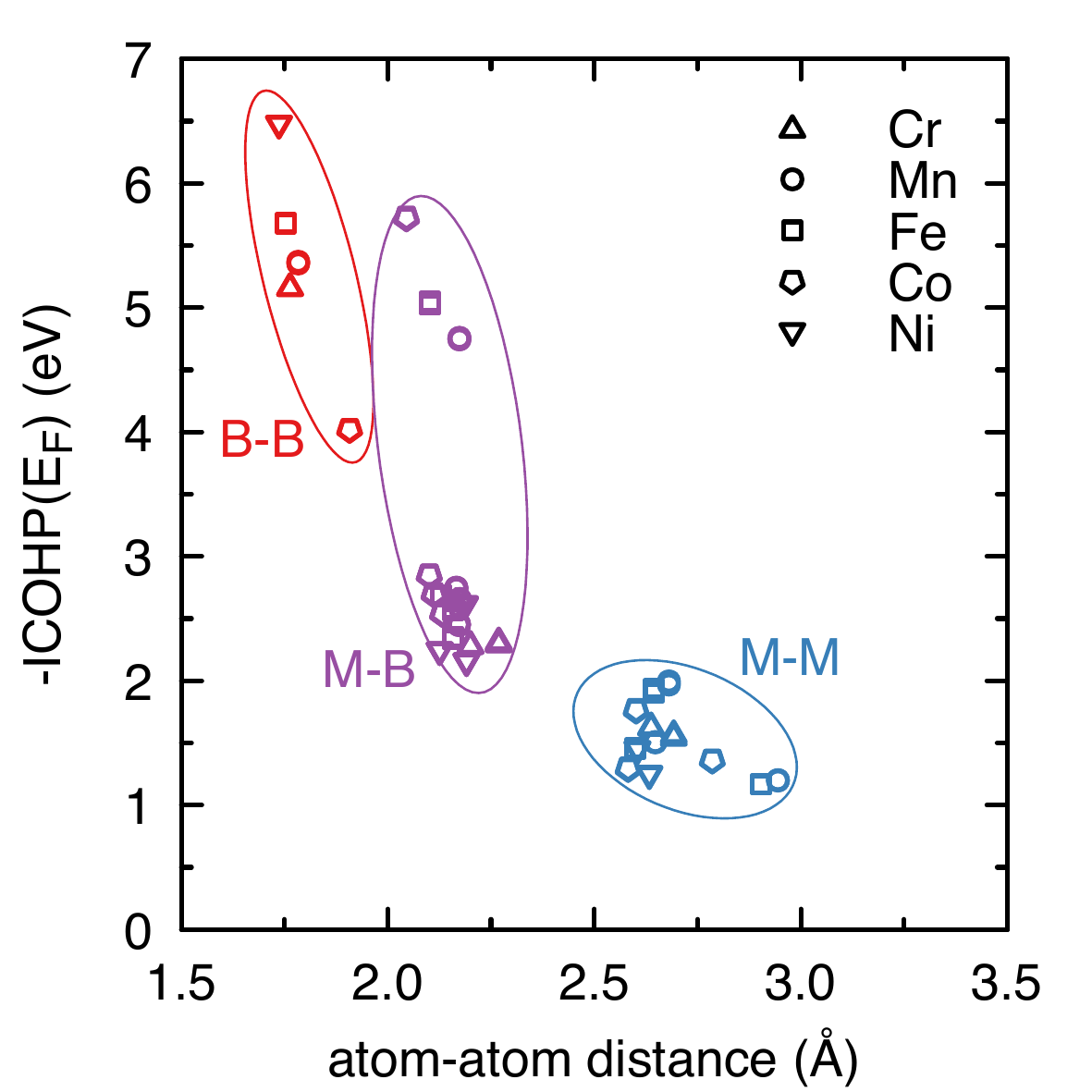}
\end{center}
\caption{\label{fig:icohp-distance-tm2alb2} 
(Color online.)
Integrated COHP up to the Fermi level, with reversed sign, for different atom-atom bonds and bond separations in (a) AlM$_2$B$_2$ and (b) MB (M = Cr, Mn, Fe, Co, Ni).}
\end{figure}

In order to discus the differences in the chemical bonding of different 
borides, in Fig. \ref{fig:icohp-distance-tm2alb2} we plotted the 
calculated integral of the 
crystal orbital Hamilton occupation (COHP), integrated up to the 
Fermi level (ICOHP) for the AlM$_2$B$_2$ and the binary 
MB compounds (M = Cr, Mn, Fe, Co, Ni), as calculated with the 
LOBSTER code \cite{Dronskowski:1993jp, Deringer:2011fg, Maintz:2013cp, Maintz:2016ee}.
In all the borides considered, the B-B bonds have a large 
values of the COHP function, signaling a high bond-strength, 
as expected for covalent bonds. 
The M-M bonds are found to have a much 
lower values of COHP, which is due to a reduced wavefunction 
overlap between orbitals centered on these atoms, and a  
reduced bond-strength. This is consistent with what is known about 
the chemical bonding of e.g. transition metal carbides \cite{price89}.
The M-B bond has 
intermediate values of COHP function, suggesting also an intermediate
bond strength.

A clear separation between the B-B bond strengths and the other 
bonds are observed 
not only for the AlM$_2$B$_2$ systems, but also for other borides, e.g. 
the monoborides 
(Fig. \ref{fig:icohp-distance-tm2alb2}b).
However, the integrated ICOHP of the binary compounds show much 
larger variations for the B-B and M-B bonds, compared to the AlM2B2 systems,
which is possibly due to 
the fact that the bonds for a three-dimensional network,
with varying crystal structures, as is 
the case 
of the binary borides, while a quasi-two-dimensional network is formed 
in the Al-borides. 
The Mn-B, Fe-B, and Co-B bond strengths in the binary borides stand out, 
which can be explained by the fact that these compounds crystallize in 
a different structure compared to CrB and NiB.

\begin{figure}[h]
(a)\includegraphics[width=0.23\textwidth]{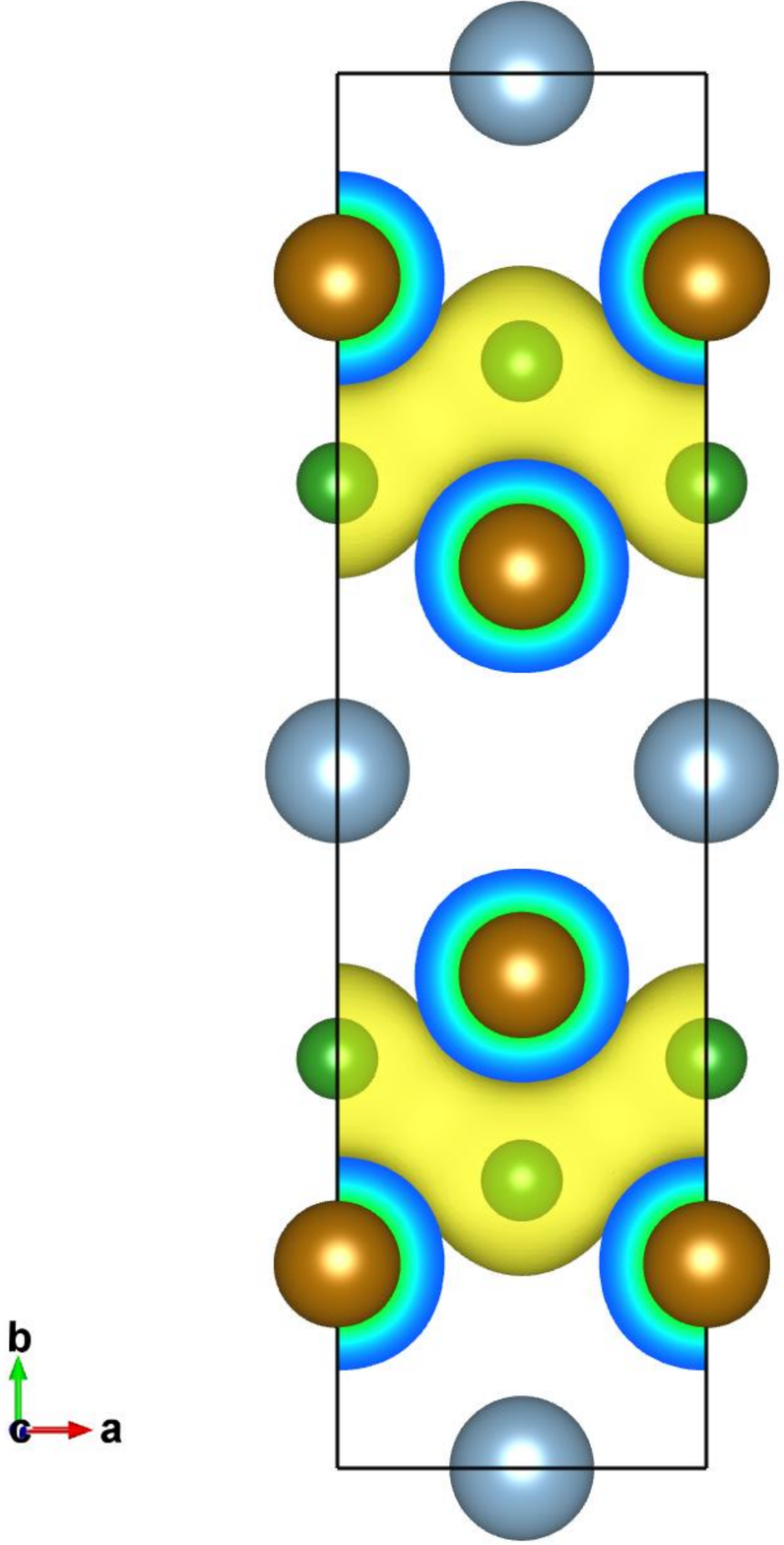}
(b)\includegraphics[width=0.70\textwidth]{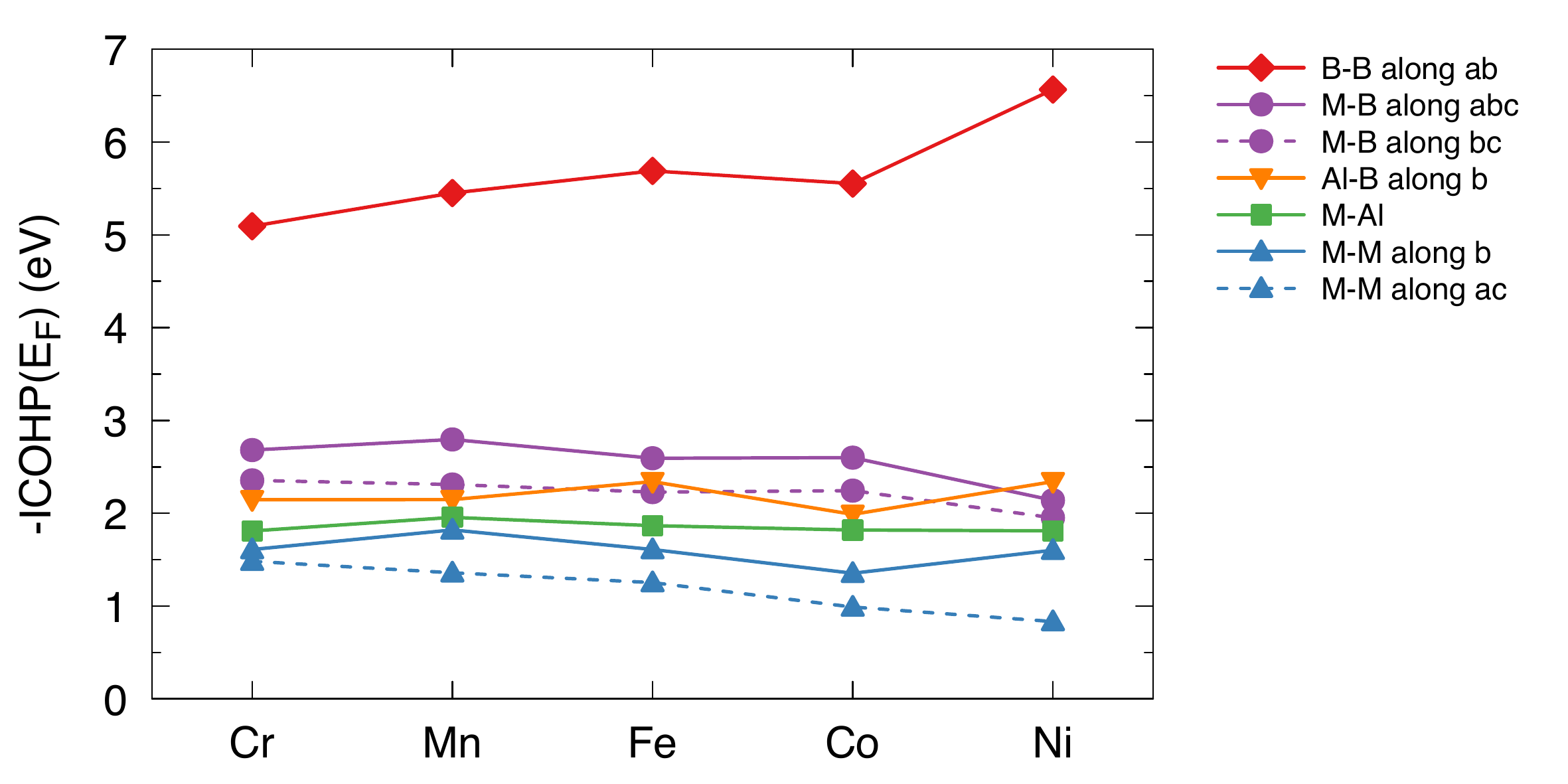}
\caption{\label{fig:icohp-bonds-tm2alb2} 
(Color online.)
a). Isosurface of the charge density of the AlFe$_2$B$_2$. 
The value of the isosurface was set to 0.09 electrons/$\rm \AA^3$. 
Similar pictures are obtained for the other AlM$_2$B$_2$ compounds. 
b) Integrated COHP up to the Fermi level, with reversed sign, 
for different bonds in AlM$_2$B$_2$.
}
\end{figure}

The quasi-two-dimensional character of the crystal structure of the 
Al-borides is evident from Fig. \ref{fig:icohp-bonds-tm2alb2}a, 
which shows an isosurface of the AlFe$_2$B$_2$ charge density. 
The largest values of the charge density are within the MB planes, 
while much lower values are obtained in the Al planes. 

The integrated COHP for the AlM$_2$B$_2$ series (M = Cr, Mn, Fe, Co, Ni) 
is shown in Fig. \ref{fig:icohp-bonds-tm2alb2}b. 
The B-B bond shows a rather steady increase with increasing $Z$, 
while the metal-metal bond strength shows a decrease. 
The latter may be explained by a filling of the antibonding states as 
the \textit{3d} occupation increases along the series, as observed 
from the COHP calculations.
This is in line with Chai et al.'s results \cite{chai2015}.

\subsection{Work of separation}
\label{wsep}

\begin{figure}[h!]
\begin{center}
\includegraphics[width=0.5\textwidth]{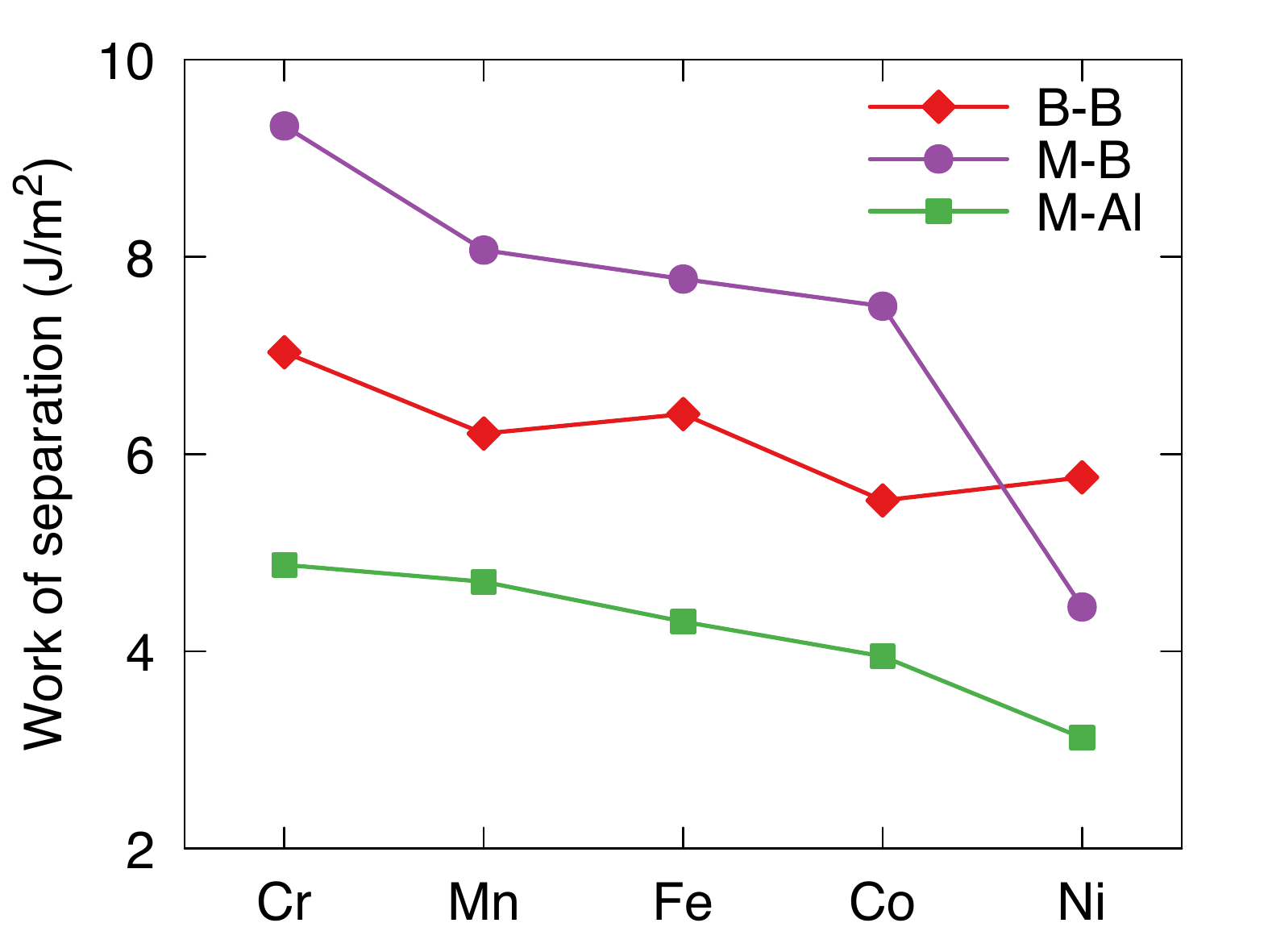}
\end{center}
\caption{\label{fig:breakbonds-tm2alb2} 
(Color online.)
Work of separation in AlM$_2$B$_2$ calculated upon separating the crystals 
between two B planes (diamonds), between M and B planes (circles), and
between M and Al planes (squares).
}
\end{figure}

In order to gain insight into the backgound of delamination,
we calculated the work of separation ($W$) in AlM$_2$B$_2$
crystals.
We separated the crystals (1) between transition metal and aluminium layers,
(2) between two boron layers within the M-B slab, 
and (3) between transition metal and boron layers.
As Fig. \ref{fig:breakbonds-tm2alb2} shows, the work of separation
decreases with increasing $Z$.
For all AlM$_2$B$_2$ borides, it is energetically most favorable 
to separate the crystals between the transition metal and aluminium layers,
i.e. to break M-Al bonds (four bonds per unit cell), 
which are weaker than M-B or B-B bonds, 
as we previously shown in Fig. \ref{fig:icohp-bonds-tm2alb2}b.
The highest work of separation corresponds to breaking the bonds
between transition metal and boron layers,
except for AlNi$_2$B$_2$, where
we obtained the highest energy for separating the crystal between 
two boron layers.
This is due to the fact that upon separation between M and B layers,
six M-B bonds per unit cell are broken, while
separation between two boron layers involves breaking of
two B-B bonds per unit cell.
AlNi$_2$B$_2$ has the strongest B-B bonds of all AlM$_2$B$_2$ borides
(see Fig. \ref{fig:icohp-bonds-tm2alb2}b), in fact, here the B-B bonds are
more than three times as strong as the M-B bonds,
therefore in this crystal separation between two boron planes requires
the highest energy.
We have performed a Bader charge analysis of 
these compounds, and find that the Al atoms have donated electrons, 
and are therefore positively charged. Hence, there is an ionic contribution 
to the weaker bond of these atoms.

On final thing should be noted with Fig. \ref{fig:breakbonds-tm2alb2}, 
namely that the trend of the bond strength shows a decreasing behavior 
of the M-B and B-B bond, as the series is traversed. 
This behavior is not reflected in the trend of 
Fig. \ref{fig:icohp-bonds-tm2alb2}b, where the B-B ICOHP curve is 
increasing along the series. It should be noted that the
ICOHP is by no means an exact measure of the chemical binding, 
it is merely an indicator, while the data in 
Fig. \ref{fig:breakbonds-tm2alb2} shows total energy differences, and 
are more reliable. 

\subsection{Experimental results}
\label{expres}

Samples with a nominal composition AlM$_2$B$_2$ (M=Cr, Mn, Fe, Co, Ni) 
were synthesized by arc-melting followed by annealing 
(see Section \ref{exp}).  
X-ray diffraction (XRD) revealed that AlCr$_2$B$_2$, AlMn$_2$B$_2$, and AlFe$_2$B$_2$
were formed.  
The unit cell parameters are summarized in Table \ref{tab_theor_cellpar} 
and show  a good agreement with calculations, 
as discussed in Section \ref{str}.
For the experimental synthesis of 
AlCo$_2$B$_2$, a multiphase sample was observed from the XRD measurements, 
including an unknown phase. 
From SEM  (not shown), a ternary phase is observed with a similar 
composition to the expected AlCo$_2$B$_2$ 
(Al 23 at.\%, Co 34 at.\%, B 43 at.\%, measured with SEM-EDS). 
The unit cell of the unknown phase can be indexed from single crystal diffraction
as monoclinic with the cell parameters
$a$=2.924(4) \AA, $b$=6.107(8) \AA, $c$=8.522(8) \AA,
and $\beta$=76.37(9)$^{\circ}$.
Unfortunately, due to strong twinning in the sample, we have not been able
to solve the complete crystal structure although measurements with single
crystal XRD as well as electron diffraction and EDS measurements have been
performed.
This monoclinic phase decomposes after heat treatments
(900 $^{\circ}$C for 14 d, followed by quenching in cold water),
suggesting metastability.
For the AlNi$_2$B$_2$ sample, no ternary phases similar to AlM$_2$B$_2$
were observed.
Consequently, our experimental studies confirm the theoretical results 
in Fig. \ref{fig_deltae} suggesting only Cr, Mn and Fe form stable 
AlM$_2$B$_2$ borides.  

The Vickers hardness of the AlM$_2$B$_2$ phases was determined to be
10.4(3) GPa for AlCr$_2$B$_2$, 7.3(3) GPa for AlMn$_2$B$_2$, and 9.5(3) GPa for 
AlFe$_2$B$_2$. 
This is in line with recently reported values of
6.0-7.0 GPa for AlCr$_2$B$_2$,
7.0-9.6 GPa for AlMn$_2$B$_2$, and
11.6-14.7 GPa for AlFe$_2$B$_2$ \cite{ade2015}.
These hardness values are significantly lower than those of transition 
metal mono- and diborides, which have a Vickers hardness around 20-25 GPa,
and also lower than the hardness values estimated from the 
theoretical results ($>$17 GPa) using the models in Ref. \cite{ivanovskii2013}.
As will be discussed below we attribute this deviation is due to a 
nanolaminated structure similar to the well-known MAX-phases.

\begin{figure}[t!]
\begin{center}
(a)\includegraphics[width=7cm]{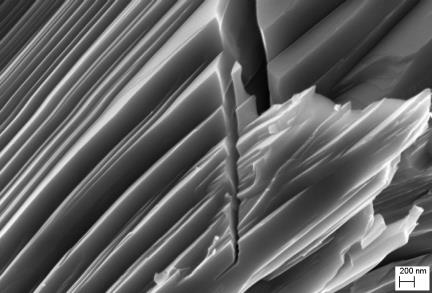}
(b)\includegraphics[width=7cm]{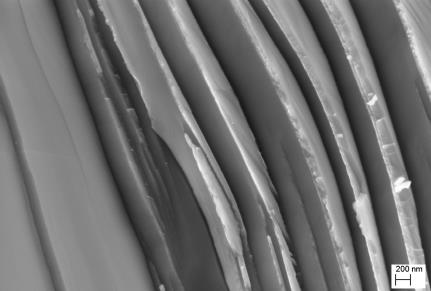}
\end{center}
\caption{
Delamination (a) in AlCr$_2$B$_2$ and (b) in AlFe$_2$B$_2$ 
(SEM micrographs).
}
\label{fig_expt}
\end{figure}

We also studied the deformation of AlCr$_2$B$_2$ and AlFe$_2$B$_2$ 
in SEM by pressing a sharp tool into the surface of a polished sample. 
SEM micrographs of the damaged areas show a clear laminated fracture 
behavior 
(see Fig. \ref{fig_expt}).
A clear difference in deformation behavior was observed for AlCr$_2$B$_2$ 
and AlFe$_2$B$_2$.
AlCr$_2$B$_2$ has more cracks and thinner flakes than AlFe$_2$B$_2$.
The AlMn$_2$B$_2$ sample showed similar delamination phenomena as AlFe2B2,
but to a lesser extent due to a less well sintered sample, and is not shown here.

\section{Discussion of the nanolaminate behavior of AlM$_2$B$_2$}
\label{disc}

The theoretical results clearly show that AlM$_2$B$_2$ ternary borides 
represent nanolaminated systems.
These borides have a layered structure where M-B slabs are separated 
by Al layers.
The M-B slabs incorporate strong covalent B-B and M-B bonds, 
which gives a very strong cohesion to these slabs.
The interlayer Al-B and M-B bonds, however, are much weaker,
compared to the sum of B-B and M-B bonds within the M-B slabs.
This makes delamination possible.
All AlM$_2$B$_2$ compounds have a significantly larger electron density
in the M-B slabs than in the Al layers and between Al layers and M-B slabs.
Thus, delamination is expected to occur between M-B
slabs and Al layers in all ternary borides.
This is supported by the calculated work of separation in these borides.
We found that all AlM$_2$B$_2$ crystals are easiest to separate 
between the transition metal and aluminium layers.
For the stable AlM$_2$B$_2$ borides, 
this requires about 2 J/m$^2$ less energy
than separation between two boron layers, and about 4 J/m$^2$ less energy
than separation between the transition metal and boron layers.

Theoretical models estimating hardness from elastic constants
\cite{ivanovskii2013},
which are able to roughly
reproduce hardness of e.g. transition metal diborides,
predict significantly higher hardness (above 17 GPa) 
for the stable AlM$_2$B$_2$ borides, 
than the measured Vickers hardness, which is
10.4(3) GPa for AlCr$_2$B$_2$,
7.3(3) GPa for AlMn$_2$B$_2$, and
9.5(3) GPa for AlFe$_2$B$_2$.
Furthermore, the measured Vickers hardness of AlM$_2$B$_2$
is significantly 
lower than those of transition metal mono- and diborides.
These differences between calculated and measured hardness, as well
as between the measured hardness of ternary AlM$_2$B$_2$ borides
and transition metal mono- and diborides
are due to the ability of AlM$_2$B$_2$ borides for delamination.
This is similar to MAX phases, where the nanolaminated MAX phases
have a reduced Vickers hardness of around 5-6 GPa, compared to
the 20-30 GPa hardness of transition metal carbides.
In contrast to the MAX-phases no kink or shear band deformation 
can be observed.

Finally, we presented experimental evidence of delamination 
in AlCr$_2$B$_2$ and AlFe$_2$B$_2$
(see Fig. \ref{fig_expt}).
We found more cracks in AlCr$_2$B$_2$ (Fig. \ref{fig_expt}a)
than in AlFe$_2$B$_2$ (Fig. \ref{fig_expt}b). 
This is in agreement with the theoretical elastic constants,
namely that AlCr$_2$B$_2$ has higher $c_{11}$ (552.2 GPa) than 
AlFe$_2$B$_2$ with a $c_{11}$ of 447.0 GPa, i.e. to distort
the AlCr$_2$B$_2$ crystal along the crystallographic axis $a$, requires
more energy than in AlFe$_2$B$_2$. 
This in turn can lead to easier crack formation in AlCr$_2$B$_2$.
Figure \ref{fig_expt} also shows that there are more flakes 
in the delaminated AlCr$_2$B$_2$ sample than in AlFe$_2$B$_2$,
suggesting that it should be easier to delaminate AlCr$_2$B$_2$.
This is in line with the theoretically
predicted
M-Al bond strengths, namely 
we obtained a lower -ICOHP for the Cr-Al than the Fe-Al bonds, suggesting
that the Cr-Al bond is weaker than the Fe-Al bond.
However,
we calculated slightly higher work of separation for AlCr$_2$B$_2$
than for AlFe$_2$B$_2$ (Fig. \ref{fig:icohp-bonds-tm2alb2}b), 
which seems to be in contradiction 
with the experimental results.
We speculate that in order to predict the energetics of delamination, 
the additional factor of chemical bonding has to be included. 
When delamination occurs, the freshly exposed surfaces will rapidly 
adsorb e.g. oxygen forming metal-oxygen bonds and eventually surface oxides. 
This is an exothermic reaction and the total delamination process 
will therefore be influenced by the sum of the energy cost to break 
M-Al bonds and energy gain upon forming surface oxides.
In fact, the enthalpy of formation of e.g. 
Cr$_2$O$_3$ (-1140 kJ/mol) is lower than that of 
Fe$_2$O$_3$ (-824 kJ/mol), 
therefore the energy gain to form Cr$_2$O$_3$ would be larger.
Thus, it could be energetically more favorable to delaminate AlCr$_2$B$_2$.

\section{Conclusions}
\label{concl}

In this study we have
examined the phase stability and elastic properties of ternary borides, 
namely AlCr$_2$B$_2$, AlMn$_2$B$_2$, AlFe$_2$B$_2$, AlCo$_2$B$_2$, and AlNi$_2$B$_2$.
We calculated ther unit cell parameters and found a good agreement between theoretical and 
measured experimental cell parameters.
The phase stability of the borides has also been investigated theoretically, 
and we find that 
AlCr$_2$B$_2$, AlMn$_2$B$_2$, and AlFe$_2$B$_2$ are stable,
while AlCo$_2$B$_2$, and AlNi$_2$B$_2$ are metastable phases.
We calculated the elastic properties of all compounds, as well,
and found that all the borides are
mechanically stable, as their elastic constants fulfill the stability criteria.
The bulk modulus to shear modulus ratio, $B/G$,
increases from AlCr$_2$B$_2$ to AlNi$_2$B$_2$,
i.e. the borides become more ductile for the heavier compounds.
The first three borides in the series, 
AlCr$_2$B$_2$, AlMn$_2$B$_2$, and AlFe$_2$B$_2$ are more brittle, while
AlCo$_2$B$_2$ and AlNi$_2$B$_2$ are more ductile.
We proposed that the mechanical properties of AlFe$_2$B$_2$ could be improved, 
namely shifted towards ductile behavior, by alloying it with 
cobalt or nickel, or a combination of them.
Our results suggest that 
Al(Fe$_x$Co$_{1-x}$)$_2$B$_2$,
Al(Fe$_x$Ni$_{1-x}$)$_2$B$_2$, or even
Al(Fe$_x$Co$_y$Ni$_{1-x-y}$)$_2$B$_2$
could be stable, and at the same time 
have improved elasticity compared to pure AlFe$_2$B$_2$.

The chemical bonding in these layered ternary borides
was investigated theoretically by several tools, and we found
strong covalent B-B and M-B bonds in the M-B layers, and weaker
bonds between M-B and Al layers.
Accordingly, all AlM$_2$B$_2$ borides 
have high electron density in the M-B slabs, 
and significantly lower density in the Al layers 
and between Al layers and M-B slabs.
The chemical bonding of Al in these compounds have a significant ionic component.
We also calculated the work of separation between B-B, M-B and M-Al
planes, and found that 
all AlM$_2$B$_2$ crystals are easiest to separate 
between the transition metal and aluminium layers, 
and we have predicted that delamination
occurs between
the M-B slabs and the Al layers.

The Vickers hardness we detected for the experimental samples, 
varies between
7.3 and 10.4 GPa, which is about 30\% of the hardness of
transition metal mono- and diborides. 
This significant reduction in hardness is 
due to their nanolaminated structure.
Finally,
applying SEM measurements, 
we demonstrated that AlCr$_2$B$_2$ and AlFe$_2$B$_2$
indeed 
show a laminated fracture behavior.

\vspace{0.6cm}

{\bf Acknowledgements}

The Swedish Research Council is acknowledged for financial support. 
O.E. also acknowledges
financial support from VR, eSSENCE and the KAW 
(projects 2013.0020 and 2012.0031).
We also acknowledge the Swedish
National Infrastructure for Computing (SNIC) for the allocation
of time in high performance supercomputers at NSC, HPC2N and PDC.

\clearpage

\begin{table}[ht]
\begin{center}
\caption{
Theoretical and experimental unit cell parameters (in \AA) 
and volumes (in \AA$^3$)
of AlM$_2$B$_2$ (M=Cr, Mn, Fe, Co, Ni).
}
\label{tab_theor_cellpar}
\begin{tabular}{ccccccccc}
\hline
Substance & $a_{\rm theor}$ & $b_{\rm theor}$ & $c_{\rm theor}$ & $V_{\rm theor}$ & $a_{\rm expt}$ & $b_{\rm expt}$ & $c_{\rm expt}$ & $V_{\rm expt}$ \\
\hline
AlCr$_2$B$_2$ & 2.9232 & 11.0511 & 2.9334 & 94.7622 & 2.9387(3) & 11.0605(11) & 2.9714(3) & 96.58(2) \\
AlMn$_2$B$_2$ & 2.8949 & 11.0750 & 2.8306 & 90.7519 & 2.9231(2) & 11.0698(9)  & 2.8993(2) & 93.82(1) \\
AlFe$_2$B$_2$ & 2.9162 & 11.0225 & 2.8515 & 91.6581 & 2.9258(4) & 11.0278(9)  & 2.8658(3) & 92.46(2) \\
AlCo$_2$B$_2$ & 2.9659 & 11.3303 & 2.6833 & 90.1711 &           &             &           &         \\
AlNi$_2$B$_2$ & 2.9779 & 11.0403 & 2.8497 & 93.6893 &           &             &           &         \\
\hline
\end{tabular}
\end{center}
\end{table}

\begin{table}[ht]
\begin{center}
\caption{
Theoretical single crystal elastic constants ($c_{ij}$ in GPa)
of AlM$_2$B$_2$ (M=Cr, Mn, Fe, Co, Ni).
}
\label{tab_cij}
\begin{tabular}{cccccccccc}
\hline
Substance & $c_{11}$ & $c_{22}$ & $c_{33}$ & $c_{44}$ & $c_{55}$ & $c_{66}$ & $c_{12}$ & $c_{13}$ & $c_{23}$ \\
\hline
AlCr$_2$B$_2$ & 552.2 & 495.7 & 478.6 & 198.3 & 221.5 & 195.8 & 124.0 & 133.5 & 147.3 \\
AlMn$_2$B$_2$ & 486.0 & 413.1 & 478.4 & 152.2 & 192.2 & 186.9 & 193.5 & 132.1 & 140.0 \\
AlFe$_2$B$_2$ & 447.0 & 402.7 & 334.6 & 140.2 & 166.3 & 156.2 & 170.1 & 133.9 & 156.4 \\
AlCo$_2$B$_2$ & 380.2 & 317.1 & 319.9 & 127.6 & 153.1 & 102.2 & 193.3 & 199.0 & 186.9 \\
AlNi$_2$B$_2$ & 375.5 & 332.7 & 253.4 & 110.2 & 101.4 & 118.7 & 154.8 & 168.6 & 178.9 \\
\hline
\end{tabular}
\end{center}
\end{table}

\begin{table}[ht]
\begin{center}
\caption{
Theoretical bulk moduli ($B$ in GPa), shear moduli ($G$ in GPa),
Young's moduli ($E$ in GPa), Poisson ratios ($\nu$), and 
$B$/$G$ ratios
of AlM$_2$B$_2$ (M=Cr, Mn, Fe, Co, Ni).
}
\label{tab_polycr_elastic_const}
\begin{tabular}{ccccccc}
\hline
Substance & B & G & E & $\nu$ & B/G &  \\
\hline
AlCr$_2$B$_2$ & 259.6 & 197.9 & 473.4 & 0.20 & 1.31 & Voight \\
              & 259.2 & 196.3 & 470.2 & 0.20 & 1.32 & Reuss  \\
              & 259.4 & 197.1 & 471.8 & 0.20 & 1.32 & Hill   \\ \hline
AlMn$_2$B$_2$ & 256.5 & 167.0 & 411.8 & 0.23 & 1.54 & Voight \\
              & 255.5 & 162.6 & 402.5 & 0.24 & 1.57 & Reuss  \\
              & 256.0 & 164.8 & 407.1 & 0.23 & 1.55 & Hill   \\ \hline
AlFe$_2$B$_2$ & 233.9 & 140.8 & 351.8 & 0.25 & 1.66 & Voight \\
              & 229.6 & 136.9 & 342.5 & 0.25 & 1.68 & Reuss  \\
              & 231.8 & 138.8 & 347.2 & 0.25 & 1.67 & Hill   \\ \hline
AlCo$_2$B$_2$ & 241.7 & 105.8 & 277.0 & 0.31 & 2.28 & Voight \\
              & 239.4 &  96.3 & 254.7 & 0.32 & 2.49 & Reuss  \\
              & 240.5 & 101.0 & 265.9 & 0.32 & 2.38 & Hill   \\ \hline
AlNi$_2$B$_2$ & 218.5 & 96.7 &  252.8 & 0.31 & 2.26 & Voight \\
              & 213.8 & 87.8 &  231.7 & 0.32 & 2.43 & Reuss  \\
              & 216.1 & 92.2 &  242.3 & 0.31 & 2.34 & Hill   \\
\hline
\end{tabular}
\end{center}
\end{table}

\begin{table}[ht]
\begin{center}
\caption{
Theoretical Cauchy pressures ($P_{\rm a}$, $P_{\rm b}$, $P_{\rm c}$ in GPa)
of AlM$_2$B$_2$ (M=Cr, Mn, Fe, Co, Ni).
}
\label{tab_cauchy}
\begin{tabular}{lccccc}
\hline
 & AlCr$_2$B$_2$ & AlMn$_2$B$_2$ & AlFe$_2$B$_2$ & AlCo$_2$B$_2$ & AlNi$_2$B$_2$ \\
\hline
$P_{\rm a}$    & -50.99   & -12.26   &  16.19   & 59.24   & 68.68   \\
$P_{\rm b}$    & -87.99   & -60.05   & -32.41   & 45.86   & 67.14   \\
$P_{\rm c}$    & -71.71   &   6.59   &  13.92   & 91.09   & 36.12   \\
\hline
\end{tabular}
\end{center}
\end{table}

\begin{table}[ht]
\begin{center}
\caption{
Theoretical shear anisotropic factors ($A_1$, $A_2$, $A_3$),
anisotropy in compressibility ($A_{\rm B}$, in \%),
anisotropy in shear ($A_{\rm G}$, in \%), and
universal elastic anisotropy index ($A_{\rm u}$)
of AlM$_2$B$_2$ (M=Cr, Mn, Fe, Co, Ni).
}
\label{tab_anisotr}
\begin{tabular}{lccccc}
\hline
 & AlCr$_2$B$_2$ & AlMn$_2$B$_2$ & AlFe$_2$B$_2$ & AlCo$_2$B$_2$ & AlNi$_2$B$_2$ \\
\hline
$A_1$       & 1.04 & 0.87 & 1.09 & 1.69 & 1.51 \\
$A_2$       & 1.30 & 1.26 & 1.57 & 2.33 & 1.78 \\
$A_3$       & 0.98 & 1.46 & 1.23 & 1.32 & 1.19 \\
$A_{\rm B}$ & 0.08 & 0.20 & 0.93 & 0.49 & 1.09 \\
$A_{\rm G}$ & 0.40 & 1.34 & 1.42 & 4.72 & 4.82 \\
$A_{\rm u}$ & 0.04 & 0.14 & 0.16 & 0.50 & 0.53 \\
\hline
\end{tabular}
\end{center}
\end{table}

\begin{table}[ht]
\begin{center}
\caption{\label{tab:magmom1} 
Calculated magnetic moments for the AlM$_2$B$_2$ compounds using 
FP-LMTO (RSPt) and PAW (VASP).
}
\begin{tabular}{ccccc}
\hline
System & $\mu$/M (RSPt) & $\mu$/f.u. (RSPt) & $\mu$/M (VASP) & $\mu$/f.u. (VASP) \\
\hline
AlMn$_2$B$_2$ & 0.39 & 0.75 & 0.43 & 0.82 \\
AlFe$_2$B$_2$ & 1.35 & 2.67 & 1.37 & 2.70 \\
AlCo$_2$B$_2$ & 0.19 & 0.39 & 0.15 & 0.31 \\
\hline
\end{tabular}
\end{center}
\end{table}

\clearpage


\begin{thebibliography}{47}
\expandafter\ifx\csname natexlab\endcsname\relax\def\natexlab#1{#1}\fi
\providecommand{\url}[1]{\texttt{#1}}
\providecommand{\href}[2]{#2}
\providecommand{\path}[1]{#1}
\providecommand{\DOIprefix}{doi:}
\providecommand{\ArXivprefix}{arXiv:}
\providecommand{\URLprefix}{URL: }
\providecommand{\Pubmedprefix}{pmid:}
\providecommand{\doi}[1]{\href{http://dx.doi.org/#1}{\path{#1}}}
\providecommand{\Pubmed}[1]{\href{pmid:#1}{\path{#1}}}
\providecommand{\bibinfo}[2]{#2}
\ifx\xfnm\relax \def\xfnm[#1]{\unskip,\space#1}\fi
\bibitem[{Barsoum(2000)}]{barsoum2000}
\bibinfo{author}{M.~Barsoum},
\newblock \bibinfo{title}{The m$_{N+1}$ax$_n$ phases: A new class of solids;
  thermodynamically stable nanolaminates},
\newblock \bibinfo{journal}{Prog. Solid State Chem.} \bibinfo{volume}{28}
  (\bibinfo{year}{2000}) \bibinfo{pages}{201--281}.
\bibitem[{Telle et~al.(2006)Telle, Momozawa, Music, and Schneider}]{telle2006}
\bibinfo{author}{R.~Telle}, \bibinfo{author}{A.~Momozawa},
  \bibinfo{author}{D.~Music}, \bibinfo{author}{J.~Schneider},
\newblock \bibinfo{title}{Boride-based nano-laminates with max-phase-like
  behaviour},
\newblock \bibinfo{journal}{J. Solid State Chem.} \bibinfo{volume}{179}
  (\bibinfo{year}{2006}) \bibinfo{pages}{2850--2857}.
\bibitem[{Ade and Hillebrecht(2015)}]{ade2015}
\bibinfo{author}{M.~Ade}, \bibinfo{author}{H.~Hillebrecht},
\newblock \bibinfo{title}{Ternary borides cr2alb2, cr3alb4, and cr4alb6: The
  first members of the series (crb2)ncral with n = 1, 2, 3 and a unifying
  concept for ternary borides as mab-phases},
\newblock \bibinfo{journal}{Inorg. Chem.} \bibinfo{volume}{54}
  (\bibinfo{year}{2015}) \bibinfo{pages}{6122--6135}.
\bibitem[{Kota et~al.(2016)Kota, Zapata-Solvas, Ly, Lu, Elkassabany, Huon, Lee,
  Hultman, May, and Barsoum}]{kota2016}
\bibinfo{author}{S.~Kota}, \bibinfo{author}{E.~Zapata-Solvas},
  \bibinfo{author}{A.~Ly}, \bibinfo{author}{J.~Lu},
  \bibinfo{author}{O.~Elkassabany}, \bibinfo{author}{A.~Huon},
  \bibinfo{author}{W.~E. Lee}, \bibinfo{author}{L.~Hultman},
  \bibinfo{author}{S.~J. May}, \bibinfo{author}{M.~W. Barsoum},
\newblock \bibinfo{title}{Synthesis and characterization of an alumina forming
  nanolaminated boride: Moalb},
\newblock \bibinfo{journal}{Sci. Rep.} \bibinfo{volume}{6}
  (\bibinfo{year}{2016}) \bibinfo{pages}{26475}.
\bibitem[{Lu et~al.(2016)Lu, Kota, Barsoum, and Hultman}]{lu2016}
\bibinfo{author}{J.~Lu}, \bibinfo{author}{S.~Kota},
  \bibinfo{author}{M.~Barsoum}, \bibinfo{author}{L.~Hultman},
\newblock \bibinfo{title}{Atomic structure and lattice defects in nanolaminated
  ternary transition metal borides},
\newblock \bibinfo{journal}{Mater. Res. Lett.}  (\bibinfo{year}{2016}).
\bibitem[{Tan et~al.(2013)Tan, Chai, Thompson, and Shatruk}]{tan2013}
\bibinfo{author}{X.~Tan}, \bibinfo{author}{P.~Chai}, \bibinfo{author}{C.~M.
  Thompson}, \bibinfo{author}{M.~Shatruk},
\newblock \bibinfo{title}{Magnetocaloric effect in alfe2b2: Toward magnetic
  refrigerants from earth-abundant elements},
\newblock \bibinfo{journal}{J. Am. Chem. Soc.} \bibinfo{volume}{135}
  (\bibinfo{year}{2013}) \bibinfo{pages}{9553--9557}.
\bibitem[{Cedervall et~al.(2016)Cedervall, Andersson, Sarkar, Delczeg-Czirjak,
  Bergqvist, Hansen, Beran, Nordblad, and Sahlberg}]{cedervall2016}
\bibinfo{author}{J.~Cedervall}, \bibinfo{author}{M.~S. Andersson},
  \bibinfo{author}{T.~Sarkar}, \bibinfo{author}{E.~K. Delczeg-Czirjak},
  \bibinfo{author}{L.~Bergqvist}, \bibinfo{author}{T.~C. Hansen},
  \bibinfo{author}{P.~Beran}, \bibinfo{author}{P.~Nordblad},
  \bibinfo{author}{M.~Sahlberg},
\newblock \bibinfo{title}{Magnetic structure of the magnetocaloric compound
  alfe2b2},
\newblock \bibinfo{journal}{Journal of Alloys and Compounds}
  \bibinfo{volume}{664} (\bibinfo{year}{2016}) \bibinfo{pages}{784--791}.
\bibitem[{Chai et~al.(2015)Chai, Stoian, Tan, Dube, and Shatruk}]{chai2015}
\bibinfo{author}{P.~Chai}, \bibinfo{author}{S.~A. Stoian},
  \bibinfo{author}{X.~Tan}, \bibinfo{author}{P.~A. Dube},
  \bibinfo{author}{M.~Shatruk},
\newblock \bibinfo{title}{Investigation of magnetic properties and electronic
  structure of layered-structure borides alt2b2 (t=fe, mn, cr) and
  alfe2–xmnxb2},
\newblock \bibinfo{journal}{J. Solid State Chem.} \bibinfo{volume}{224}
  (\bibinfo{year}{2015}) \bibinfo{pages}{52--61}.
\bibitem[{Nie et~al.(2013)Nie, Zhou, and Zhan}]{nie2013}
\bibinfo{author}{L.~Nie}, \bibinfo{author}{W.~Zhou}, \bibinfo{author}{Y.~Zhan},
\newblock \bibinfo{title}{Theoretical investigation of the al–cr–b
  orthorhombic ternary compounds},
\newblock \bibinfo{journal}{Computational and Theoretical Chemistry}
  \bibinfo{volume}{1020} (\bibinfo{year}{2013}) \bibinfo{pages}{51--56}.
\bibitem[{Cheng et~al.(2014)Cheng, Lv, Chen, and Cai}]{cheng2014}
\bibinfo{author}{Y.~Cheng}, \bibinfo{author}{Z.~Lv}, \bibinfo{author}{X.~Chen},
  \bibinfo{author}{L.~Cai},
\newblock \bibinfo{title}{Structural, electronic and elastic properties of
  alfe2b2: First-principles study},
\newblock \bibinfo{journal}{Comp. Mat. Sci.} \bibinfo{volume}{92}
  (\bibinfo{year}{2014}) \bibinfo{pages}{253--257}.
\bibitem[{Bl{\"o}chl(1994)}]{blochl94}
\bibinfo{author}{P.~E. Bl{\"o}chl},
\newblock \bibinfo{title}{{Projector augmented-wave method}}
  \bibinfo{volume}{{50}} (\bibinfo{year}{{1994}})
  \bibinfo{pages}{17953--17979}.
\bibitem[{Kresse and Joubert(1999)}]{kresse99}
\bibinfo{author}{G.~Kresse}, \bibinfo{author}{D.~Joubert},
\newblock \bibinfo{title}{From ultrasoft pseudopotentials to the projector
  augmented-wave method},
\newblock \bibinfo{journal}{Phys. Rev. B} \bibinfo{volume}{59}
  (\bibinfo{year}{1999}) \bibinfo{pages}{1758}.
\bibitem[{Kresse and Hafner(1993)}]{vasp1}
\bibinfo{author}{G.~Kresse}, \bibinfo{author}{J.~Hafner},
\newblock \bibinfo{title}{Ab initio molecular dynamics for liquid metals}
  \bibinfo{volume}{47} (\bibinfo{year}{1993}) \bibinfo{pages}{558--561}.
\bibitem[{Kresse and Furthm{\"u}ller(1996{\natexlab{a}})}]{vasp2}
\bibinfo{author}{G.~Kresse}, \bibinfo{author}{J.~Furthm{\"u}ller},
\newblock \bibinfo{title}{Efficiency of ab-initio total energy calculations for
  metals and semiconductors using a plane-wave basis set},
\newblock \bibinfo{journal}{Comp. Mat. Sci.} \bibinfo{volume}{6}
  (\bibinfo{year}{1996}{\natexlab{a}}) \bibinfo{pages}{15--50}.
\bibitem[{Kresse and Furthm{\"u}ller(1996{\natexlab{b}})}]{vasp3}
\bibinfo{author}{G.~Kresse}, \bibinfo{author}{J.~Furthm{\"u}ller},
\newblock \bibinfo{title}{Efficient iterative schemes for ab initio
  total-energy calculations using a plane-wave basis set} \bibinfo{volume}{54}
  (\bibinfo{year}{1996}{\natexlab{b}}) \bibinfo{pages}{11169--11186}.
\bibitem[{Hohenberg and Kohn(1964)}]{hohenberg64}
\bibinfo{author}{P.~Hohenberg}, \bibinfo{author}{W.~Kohn},
\newblock \bibinfo{title}{Inhomogeneous electron gas},
\newblock \bibinfo{journal}{Phys. Rev.} \bibinfo{volume}{136}
  (\bibinfo{year}{1964}) \bibinfo{pages}{B864--B871}.
\bibitem[{Kohn and Sham(1965)}]{kohn65}
\bibinfo{author}{W.~Kohn}, \bibinfo{author}{L.~Sham},
\newblock \bibinfo{title}{Self-consistent equations including exchange and
  correlation effects} \bibinfo{volume}{140} (\bibinfo{year}{1965})
  \bibinfo{pages}{A1133--A1138}.
\bibitem[{Perdew et~al.(1996)Perdew, Burke, and Ernzerhof}]{pbe}
\bibinfo{author}{J.~P. Perdew}, \bibinfo{author}{K.~Burke},
  \bibinfo{author}{M.~Ernzerhof},
\newblock \bibinfo{title}{Generalized gradient approximation made simple}
  \bibinfo{volume}{77} (\bibinfo{year}{1996}) \bibinfo{pages}{3865--3868}.
\bibitem[{Ravindran et~al.(1998)Ravindran, Fast, Korzhavyi, Johansson, Wills,
  and Eriksson}]{ravindran98}
\bibinfo{author}{P.~Ravindran}, \bibinfo{author}{L.~Fast},
  \bibinfo{author}{P.~A. Korzhavyi}, \bibinfo{author}{B.~Johansson},
  \bibinfo{author}{J.~Wills}, \bibinfo{author}{O.~Eriksson},
\newblock \bibinfo{title}{Density functional theory for calculation of elastic
  properties of orthorhombic crystals: Application to tisi$_2$},
\newblock \bibinfo{journal}{J. Appl. Phys.} \bibinfo{volume}{84}
  (\bibinfo{year}{1998}) \bibinfo{pages}{4891--4904}.
\bibitem[{Andersen(1975)}]{Andersen:1975kh}
\bibinfo{author}{O.~K. Andersen},
\newblock \bibinfo{title}{{Linear methods in band theory}},
\newblock \bibinfo{journal}{Physical Review B} \bibinfo{volume}{12}
  (\bibinfo{year}{1975}) \bibinfo{pages}{3060--3083}.
\bibitem[{Wills et~al.(2010)Wills, Eriksson, Andersson, Delin, Grechnyev, and
  Alouani}]{Wills:2010ej}
\bibinfo{author}{J.~M. Wills}, \bibinfo{author}{O.~Eriksson},
  \bibinfo{author}{P.~Andersson}, \bibinfo{author}{A.~Delin},
  \bibinfo{author}{O.~Grechnyev}, \bibinfo{author}{M.~Alouani},
  \bibinfo{title}{{Full-Potential Electronic Structure Method}}, volume
  \bibinfo{volume}{167} of \textit{\bibinfo{series}{Springer Series in
  Solid-State Sciences}}, \bibinfo{publisher}{Springer Berlin Heidelberg},
  \bibinfo{address}{Berlin, Heidelberg}, \bibinfo{year}{2010}.
\bibitem[{Dronskowski and Bloechl(1993)}]{Dronskowski:1993jp}
\bibinfo{author}{R.~Dronskowski}, \bibinfo{author}{P.~E. Bloechl},
\newblock \bibinfo{title}{{Crystal orbital Hamilton populations (COHP):
  energy-resolved visualization of chemical bonding in solids based on
  density-functional calculations}},
\newblock \bibinfo{journal}{J. Phys. Chem.} \bibinfo{volume}{97}
  (\bibinfo{year}{1993}) \bibinfo{pages}{8617--8624}.
\bibitem[{Deringer et~al.(2011)Deringer, Tchougr{\'e}eff, and
  Dronskowski}]{Deringer:2011fg}
\bibinfo{author}{V.~L. Deringer}, \bibinfo{author}{A.~L. Tchougr{\'e}eff},
  \bibinfo{author}{R.~Dronskowski},
\newblock \bibinfo{title}{{Crystal Orbital Hamilton Population (COHP) Analysis
  As Projected from Plane-Wave Basis Sets}},
\newblock \bibinfo{journal}{J. Phys. Chem. A} \bibinfo{volume}{115}
  (\bibinfo{year}{2011}) \bibinfo{pages}{5461--5466}.
\bibitem[{Maintz et~al.(2013)Maintz, Deringer, Tchougr{\'e}eff, and
  Dronskowski}]{Maintz:2013cp}
\bibinfo{author}{S.~Maintz}, \bibinfo{author}{V.~L. Deringer},
  \bibinfo{author}{A.~L. Tchougr{\'e}eff}, \bibinfo{author}{R.~Dronskowski},
\newblock \bibinfo{title}{{Analytic projection from plane-wave and PAW
  wavefunctions and application to chemical-bonding analysis in solids}},
\newblock \bibinfo{journal}{J. Comput. Chem.} \bibinfo{volume}{34}
  (\bibinfo{year}{2013}) \bibinfo{pages}{2557--2567}.
\bibitem[{Maintz et~al.(2016)Maintz, Deringer, Tchougr{\'e}eff, and
  Dronskowski}]{Maintz:2016ee}
\bibinfo{author}{S.~Maintz}, \bibinfo{author}{V.~L. Deringer},
  \bibinfo{author}{A.~L. Tchougr{\'e}eff}, \bibinfo{author}{R.~Dronskowski},
\newblock \bibinfo{title}{{LOBSTER: A tool to extract chemical bonding from
  plane-wave based DFT}},
\newblock \bibinfo{journal}{J. Comput. Chem.} \bibinfo{volume}{37}
  (\bibinfo{year}{2016}) \bibinfo{pages}{1030--1035}.
\bibitem[{ElMassalami et~al.(2011)ElMassalami, da~S.~Oliveira, and
  Takeya}]{elmassalami2011}
\bibinfo{author}{M.~ElMassalami}, \bibinfo{author}{D.~da~S.~Oliveira},
  \bibinfo{author}{H.~Takeya},
\newblock \bibinfo{title}{On the ferromagnetism of alfe2b2},
\newblock \bibinfo{journal}{J. Magn. Magn. Mat.} \bibinfo{volume}{323}
  (\bibinfo{year}{2011}) \bibinfo{pages}{2133--2136}.
\bibitem[{Holland and Redfern(1997)}]{holland1997}
\bibinfo{author}{T.~J.~B. Holland}, \bibinfo{author}{S.~A.~T. Redfern},
\newblock \bibinfo{title}{Unit cell refinement from powder diffraction data;
  the use of regression diagnostics},
\newblock \bibinfo{journal}{Miner. Mag.} \bibinfo{volume}{61}
  (\bibinfo{year}{1997}) \bibinfo{pages}{65--77}.
\bibitem[{Jeitschko(1969)}]{jeitschko69}
\bibinfo{author}{W.~Jeitschko},
\newblock \bibinfo{title}{The crystal structure of fe$_2$alb$_2$},
\newblock \bibinfo{journal}{Acta Cryst.} \bibinfo{volume}{B25}
  (\bibinfo{year}{1969}) \bibinfo{pages}{163--165}.
\bibitem[{Felten(1956)}]{felten56}
\bibinfo{author}{E.~J. Felten},
\newblock \bibinfo{title}{The preparation of aluminum diboride, alb2},
\newblock \bibinfo{journal}{J. Am. Chem. Soc.} \bibinfo{volume}{78}
  (\bibinfo{year}{1956}) \bibinfo{pages}{5977--5978}.
\bibitem[{Kiessling(1950)}]{kiessling50}
\bibinfo{author}{R.~Kiessling},
\newblock \bibinfo{title}{The borides of manganese},
\newblock \bibinfo{journal}{Acta Chem. Scand.} \bibinfo{volume}{4}
  (\bibinfo{year}{1950}) \bibinfo{pages}{146--159}.
\bibitem[{Bjurstrom(1933)}]{bjurstrom33}
\bibinfo{author}{T.~Bjurstrom},
\newblock \bibinfo{title}{R\"ontgenanalyse der systeme eisen-bor, kobalt-bor
  und nickel-bor},
\newblock \bibinfo{journal}{Ark. Kemi Mineral. Geol.} \bibinfo{volume}{11A}
  (\bibinfo{year}{1933}) \bibinfo{pages}{1--12}.
\bibitem[{Okada et~al.(1987)Okada, Atoda, and Higashi}]{okada87}
\bibinfo{author}{S.~Okada}, \bibinfo{author}{T.~Atoda},
  \bibinfo{author}{I.~Higashi},
\newblock \bibinfo{title}{Structural investigation of cr2b3, cr3b4, and crb by
  single-crystal diffractometry},
\newblock \bibinfo{journal}{J. Solid State Chem.} \bibinfo{volume}{68}
  (\bibinfo{year}{1987}) \bibinfo{pages}{61--67}.
\bibitem[{Lugscheider et~al.(1974)Lugscheider, Knotek, and
  Reimann}]{lugscheider74}
\bibinfo{author}{E.~Lugscheider}, \bibinfo{author}{O.~Knotek},
  \bibinfo{author}{H.~Reimann},
\newblock \bibinfo{title}{The ternary system nickel---chromium---boron},
\newblock \bibinfo{journal}{Monatshefte f{\"u}r Chemie / Chemical Monthly}
  \bibinfo{volume}{105} (\bibinfo{year}{1974}) \bibinfo{pages}{80--90}.
\bibitem[{Kallel(1969)}]{kallel69}
\bibinfo{author}{A.~Kallel},
\newblock \bibinfo{title}{Ordre antiferromagnétique dans les alliages
  cr2-xfexal},
\newblock \bibinfo{journal}{C.R. Seances Acad. Sci., Ser. B}
  \bibinfo{volume}{268} (\bibinfo{year}{1969}) \bibinfo{pages}{455--458}.
\bibitem[{Kontio and Coppens(1981)}]{kontio81}
\bibinfo{author}{A.~Kontio}, \bibinfo{author}{P.~Coppens},
\newblock \bibinfo{title}{{New study of the structure of MnAl6}},
\newblock \bibinfo{journal}{{Acta Cryst.}} \bibinfo{volume}{{B37}}
  (\bibinfo{year}{{1981}}) \bibinfo{pages}{{433--435}}.
\bibitem[{der Kraan and Buschow(1986)}]{vanderkraan86}
\bibinfo{author}{A.~V. der Kraan}, \bibinfo{author}{K.~Buschow},
\newblock \bibinfo{title}{The 57fe mössbauer isomer shift in intermetallic
  compounds of iron},
\newblock \bibinfo{journal}{Physica B+C} \bibinfo{volume}{138}
  (\bibinfo{year}{1986}) \bibinfo{pages}{55--62}.
\bibitem[{Ipser and Mikula(1992)}]{ipser92}
\bibinfo{author}{H.~Ipser}, \bibinfo{author}{A.~Mikula},
\newblock \bibinfo{title}{On the ternary b$_2$-phase in the al-co-ga system},
\newblock \bibinfo{journal}{Monatshefte f{\"u}r Chemie / Chemical Monthly}
  \bibinfo{volume}{123} (\bibinfo{year}{1992}) \bibinfo{pages}{509--513}.
\bibitem[{Dutchak and Chekh(1981)}]{dutchak81}
\bibinfo{author}{Y.~Dutchak}, \bibinfo{author}{V.~Chekh},
\newblock \bibinfo{title}{High temperature x-ray diffraction study of the
  lattice dynamics of the compounds alco and alni},
\newblock \bibinfo{journal}{Russ. J. Phys. Chem.} \bibinfo{volume}{55}
  (\bibinfo{year}{1981}) \bibinfo{pages}{1326--1328}.
\bibitem[{Mouhat and Coudert(2014)}]{mouhat2014}
\bibinfo{author}{F.~Mouhat}, \bibinfo{author}{F.-X. Coudert},
\newblock \bibinfo{title}{Necessary and sufficient elastic stability conditions
  in various crystal systems},
\newblock \bibinfo{journal}{Phys. Rev. B} \bibinfo{volume}{90}
  (\bibinfo{year}{2014}) \bibinfo{pages}{224104}.
\bibitem[{Voigt(1928)}]{voigt}
\bibinfo{author}{W.~Voigt}, \bibinfo{title}{Lehrbuch der Kristallphysik},
  \bibinfo{publisher}{Teubner}, \bibinfo{address}{Leipzig},
  \bibinfo{year}{1928}.
\bibitem[{Reuss(1929)}]{reuss}
\bibinfo{author}{A.~Reuss},
\newblock \bibinfo{title}{Account of the liquid limit of mixed crystals on the
  basis of the plasticity condition for single crystal},
\newblock \bibinfo{journal}{Z. Angew. Math. Mech.} \bibinfo{volume}{9}
  (\bibinfo{year}{1929}) \bibinfo{pages}{49--58}.
\bibitem[{Tvergaard and Hutchinson(1988)}]{tvergaard88}
\bibinfo{author}{V.~Tvergaard}, \bibinfo{author}{J.~Hutchinson},
\newblock \bibinfo{title}{Microcracking in ceramics induced by thermal
  expansion or elastic anisotropy},
\newblock \bibinfo{journal}{J. Am. Ceram. Soc.} \bibinfo{volume}{71}
  (\bibinfo{year}{1988}) \bibinfo{pages}{157--166}.
\bibitem[{Chung and Buessem(1968)}]{chung}
\bibinfo{author}{D.~H. Chung}, \bibinfo{author}{W.~R. Buessem},
\newblock in: \bibinfo{editor}{F.~W. Vahldiek}, \bibinfo{editor}{S.~A. Mersol}
  (Eds.), \bibinfo{booktitle}{Anisotropy in Single Crystal Refractory
  Compound}, volume~\bibinfo{volume}{2}, \bibinfo{publisher}{Plenum},
  \bibinfo{address}{New York}, \bibinfo{year}{1968}, p. \bibinfo{pages}{217}.
\bibitem[{Ranganathan and Ostoja-Starzewski(2008)}]{ranganathan2008}
\bibinfo{author}{S.~I. Ranganathan}, \bibinfo{author}{M.~Ostoja-Starzewski},
\newblock \bibinfo{title}{Universal elastic anisotropy index},
\newblock \bibinfo{journal}{Phys. Rev. Lett.} \bibinfo{volume}{101}
  (\bibinfo{year}{2008}) \bibinfo{pages}{055504}.
\bibitem[{Ivanovskii(2013)}]{ivanovskii2013}
\bibinfo{author}{A.~Ivanovskii},
\newblock \bibinfo{title}{Hardness of hexagonal alb2-like diborides of s, p and
  d metals from semi-empirical estimations},
\newblock \bibinfo{journal}{International Journal of Refractory Metals and Hard
  Materials} \bibinfo{volume}{36} (\bibinfo{year}{2013})
  \bibinfo{pages}{179--182}.
\bibitem[{Lejaeghere et~al.(2016)Lejaeghere, Bihlmayer, Bj{\"o}rkman, Blaha,
  Bl{\"u}gel, and et~al.}]{lejaeghere2016}
\bibinfo{author}{K.~Lejaeghere}, \bibinfo{author}{G.~Bihlmayer},
  \bibinfo{author}{T.~Bj{\"o}rkman}, \bibinfo{author}{P.~Blaha},
  \bibinfo{author}{S.~Bl{\"u}gel}, \bibinfo{author}{et~al.},
\newblock \bibinfo{title}{Reproducibility in density functional theory
  calculations of solids},
\newblock \bibinfo{journal}{Science} \bibinfo{volume}{351}
  (\bibinfo{year}{2016}).
\bibitem[{Price and Cooper(1989)}]{price89}
\bibinfo{author}{D.~L. Price}, \bibinfo{author}{B.~R. Cooper},
\newblock \bibinfo{title}{Total energies and bonding for crystallographic
  structures in titanium-carbon and tungsten-carbon systems},
\newblock \bibinfo{journal}{Phys. Rev. B} \bibinfo{volume}{39}
  (\bibinfo{year}{1989}) \bibinfo{pages}{4945--4957}.

\end{thebibliography}

\end{document}